\documentstyle[aaspp4,epsf,11pt]{article}
\newcommand{\xt}[1]{\mbox{$\times 10^{#1}$}}
\newcommand{\nm}{{\rm nm}}

\newcommand{\x}[1]{\hspace*{#1mm}}
\newcommand{\zm}{M$_\odot$}
\begin{document}

\title{Properties of Dust Grains in Planetary Nebulae\\
I. The Ionized Region of NGC~6445}
\author{Peter~A.~M. van Hoof\altaffilmark{1}}
\affil{University of Kentucky, Dept.\ of Physics and Astronomy,
177 CP Building,\\ Lexington, KY 40506--0055, USA}
\authoremail{peter@pa.uky.edu}
\author{Griet~C. Van de Steene}
\affil{Research School of Astronomy and Astrophysics,
Mount Stromlo Observatory,\\
Private Bag, Weston Creek, ACT 2611, Australia}
\authoremail{gsteene@mso.anu.edu.au}
\author{Douwe~A. Beintema}
\affil{Laboratory for Space Research, Postbus 800, NL-9700~AV Groningen,
The Netherlands}
\authoremail{douwe@sron.rug.nl}
\author{P.~G. Martin}
\affil{Canadian Institute for Theoretical Astrophysics, McLennan Labs,
University of Toronto,\\ 60 St.\ George Street, Toronto, ON M5S 3H8, Canada}
\authoremail{pgmartin@cita.utoronto.ca}
\author{Stuart~R. Pottasch}
\affil{Kapteyn Astronomical Institute, Postbus 800, NL-9700~AV Groningen,
The Netherlands}
\authoremail{pottasch@astro.rug.nl}
\and
\author{Gary~J. Ferland\altaffilmark{1}}
\affil{University of Kentucky, Dept.\ of Physics and Astronomy,
177 CP Building,\\ Lexington, KY 40506--0055, USA}
\authoremail{gary@cloud9.pa.uky.edu}
\altaffiltext{1}{Visiting Astronomer at the Canadian Institute for Theoretical
Astrophysics, Toronto, Canada.}

\slugcomment{Accepted for publication in the Astrophysical Journal.}

\lefthead{P.~A.~M. van Hoof et al.}
\righthead{Dust Grains in PNe -- The Ionized Region of NGC~6445}

\newpage

\begin{abstract}
One of the factors influencing the spectral evolution of a planetary nebula is
the fate of the dust grains that are emitting the infrared continuum. Several
processes have been proposed that either destroy the grains or remove them
from the ionized region. To test whether these processes are effective, we
study new infrared spectra of the evolved nebula NGC~6445. These data show
that the thermal emission from the grains is very cool and has a low flux
compared to H$\beta$. A model of the ionized region is constructed, using the
photo-ionization code {\sc cloudy} 90.05. Based on this model, we show from
depletions in the gas phase elements that little grain destruction can have
occurred in the ionized region of NGC~6445. We also argue that dust-gas
separation in the nebula is not plausible. The most likely conclusion is that
grains are residing inside the ionized region of NGC~6445 and that the low
temperature and flux of the grain emission are caused by the low luminosity of
the central star and the low optical depth of the grains. This implies that
the bulk of the silicon-bearing grains in this nebula were able to survive
exposure to hard UV photons for at least several thousands of years,
contradicting previously published results.

A comparison between optical and infrared diagnostic line ratios gives a
marginal indication for the presence of a $t^2$-effect in the nebula. However,
the evidence is not convincing and the differences could also be explained by
uncertainties in the absolute flux calibration of the spectra, the aperture
corrections that have been applied or the collisional cross sections. The
photo-ionization model allowed an accurate determination of the central star
temperature based on model atmospheres. The resulting value of 184~kK is in
good agreement with the average of all published Zanstra temperatures based on
a blackbody approximation.

The off-source spectrum taken with LWS clearly shows the presence of a warm
cirrus component with a temperature of 24~K as well as a very cold component
with a temperature of 7~K. Since our observation encompasses only a small
region of the sky, it is not clear how extended the 7~K component is and
whether it contributed significantly to the FIRAS spectrum taken by {\it
COBE}. Because our line of sight is in the galactic plane, the very cold
component could be a starless core.
\end{abstract}


\newpage

\section{Introduction}

When low- and intermediate mass stars are approaching the end of their
evolution, they go through a period of heavy mass loss known as the Asymptotic
Giant Branch (AGB) stage. In this phase the star is a red giant which consists
of a degenerate C/O core surrounded by a very tenuous envelope. During the AGB
stage the star loses a substantial fraction of its initial mass, eventually
leaving only the degenerate core of typically 0.6~$M_\odot$ to 0.8~$M_\odot$
covered with a thin layer of hydrogen. This core will eventually evolve into a
white dwarf. It is generally assumed that dust formation is very efficient in
the AGB outflow so that large amounts of dust grains are formed during that
phase. Once the envelope mass of the central star drops below a critical
value, the mass-loss rate will drop several orders of magnitude and the star
will enter the post-AGB phase of its evolution. During this phase the central
star will heat up and the AGB ejecta will expand, forming a detached shell.
Once the central star reaches a temperature of 20~kK to 30~kK, it will start
to ionize the AGB shell and a Planetary Nebula (PN) will form. After this the
central star will continue heating up at roughly constant luminosity until
finally the nuclear reactions cease and the central star will decrease both in
temperature and luminosity. All this time the detached AGB shell will continue
to expand and undergo hydrodynamic interactions with the fast wind emanating
from the central star. When the central star enters the cooling track, the
number of ionizing photons emitted by the star will decline sharply. If this
decline in luminosity is fast enough (i.e., if the central star is massive
enough), this will cause the ionized region to shrink and the outermost parts
of the nebula will start recombining. Eventually the cooling rate of the
central star will slow down and the nebula will dissolve in the Inter-Stellar
Medium (ISM), leaving a bare white dwarf. Both the evolution of the central
star and the expansion of the AGB ejecta will have a profound effect on the
infrared emission from the dust grains, causing a substantial evolution of
both the total infrared flux and the temperature of the dust grains.

During the post-AGB phase, the combined effects of increasing stellar
temperature (which will increase the grain temperature) and nebular expansion
(which will decrease the grain temperature) cannot be predicted easily and
detailed modeling is necessary. A comprehensive discussion of the evolution of
the infrared spectrum during this phase can be found in van Hoof, Oudmaijer \&
Waters\markcite{vh97a} (1997a). However, after the central star has reached a
temperature of 40~kK to 50~kK, the effect of the rising stellar temperature on
the grains will be reversed. This is caused by the fact that the peak of the
stellar spectrum shifts to ever shorter wavelengths, beyond the peak of the
grain absorption cross section, thereby making grain heating less efficient.
Therefore, once the nebula is in the PN phase, the evolution of the infrared
spectrum is clear. Both the ratio of the total infrared flux over the H$\beta$
flux and the average dust temperature will decrease with increasing
(Str\"omgren) radius of the nebula. However, a considerable amount of scatter
around this average evolution can be expected in a statistical sample, due to
the particulars of the evolution of each individual nebula.

One of the factors influencing the evolution of the infrared spectrum is the
dust grains themselves. In the previous discussion we implicitly assumed that
nothing much happened to the grains; however, this is not at all clear. It has
been proposed that the dust grains could be destroyed as the PN evolves (e.g.,
Natta \& Panagia\markcite{np81} 1981; Pottasch\markcite{po87} 1987), either
due to spallation by hard UV photons or shocks. It is known that UV photons
are abundantly present in PNe, as is the case for shocks which are formed due
to the interaction of the fast wind with the slow AGB ejecta. Hence, if these
processes are efficient, the ionized regions of evolved PNe should be free of
dust grains. Another possibility is that dust grains are separated from the
gas due to radiation pressure. This could lead to a general flow of the grains
with respect to the gas or, if instabilities occur, to the formation of small,
high-density globules which would be neutral or partially ionized. Such
globules have been observed, suggesting that at least in some nebulae this
mechanism is effective.

\begin{figure}
\mbox{\epsfxsize=0.6\textwidth\epsfbox[54 360 528 719]{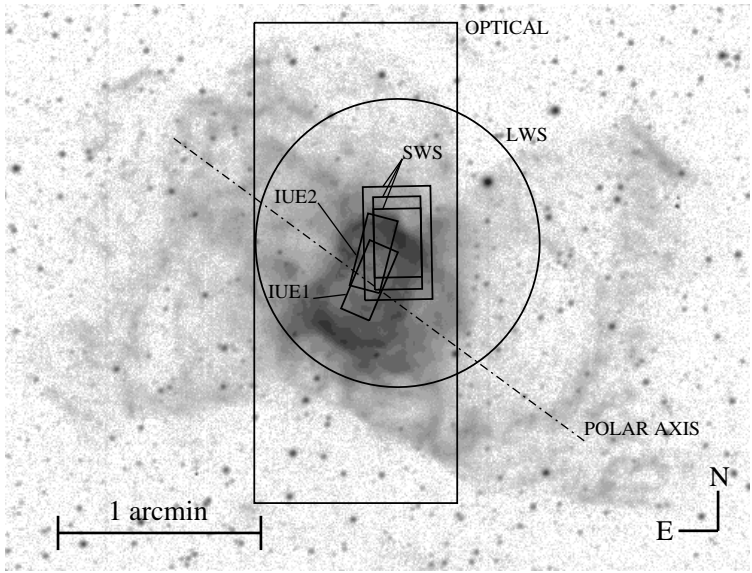}}
\figcaption[f1.ps]
{The various apertures over-plotted on an H$\alpha$+[\ion{N}{2}] image of
NGC~6445 (Schwarz et al.\protect\markcite{sc92} 1992).
The intensity has been scaled logarithmically to emphasize the low
surface brightness regions.
\label{aperture}}
\end{figure}

\begin{figure}
\mbox{\epsfxsize=0.8\textwidth\epsfbox[18 314 535 592]{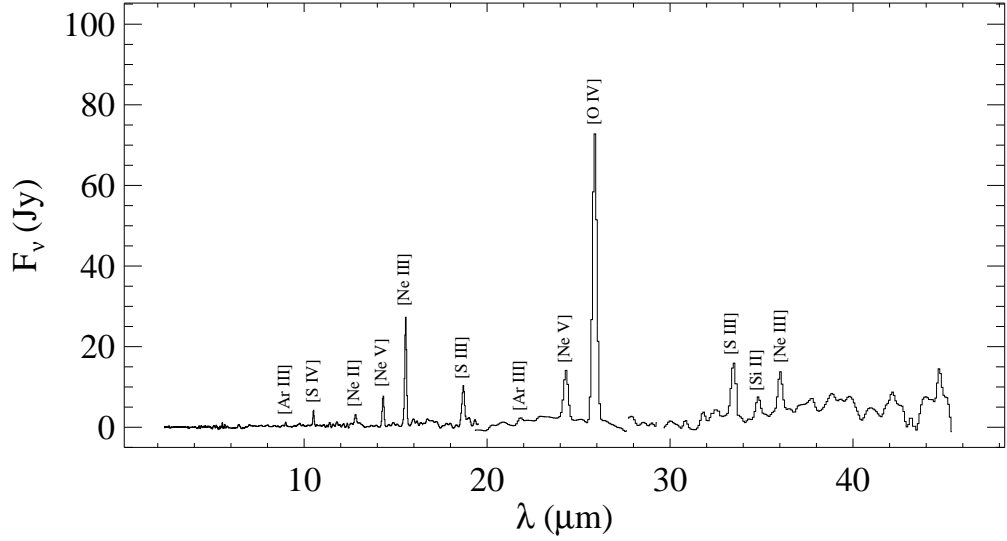}}
\figcaption[f2.ps]
{The {\it ISO}-SWS spectrum of the planetary nebula NGC~6445. \label{sws:spec}}
\end{figure}

To investigate these points in more detail, we shall analyze the infrared
spectrum of an evolved PN and search for indications of any of the processes
we mentioned above. This nebula is NGC~6445 and it is classified as a bipolar
type~I PN (Corradi \& Schwarz\markcite{cs95} 1995; Perinotto\markcite{pe91}
1991 respectively), a class assumed to originate from massive progenitors.
Based on its low density, the nebula appears well evolved. Both the ionized
and neutral material (CO) show a high expansion velocity of 38~km\,s$^{-1}$
(Weinberger\markcite{we89} 1989) and 33~km\,s$^{-1}$ (Huggins \&
Healy\markcite{hh89} 1989) respectively. These properties are consistent with
a central star which was luminous in the early stages of its post-AGB
evolution and therefore of high mass. Also the high central star temperature
we find (see \S~\ref{model}) is consistent with this. An image of the nebula
is shown in Figure~\ref{aperture}. We have obtained an SWS spectrum of this
nebula, which is remarkable in the fact that it shows virtually no dust
emission (see Fig.~\ref{sws:spec}). This is indicative of a low $F({\rm
IR})/F({\rm H}\beta)$ ratio and dust temperature for this object, which
confirms the evolved status of this PN. It also makes the object very suitable
to investigate the points mentioned earlier.

\placefigure{aperture}

First we will discuss the observations gathered for this source in
\S~\ref{obs:sec}. Since the nebula is very large, aperture corrections are
needed. These are discussed in \S~\ref{aper:corr}. Next we discuss the
reddening correction of the spectra in \S~\ref{reddening}. Some details
concerning the observed spectrum, including the possible influence of a
recombining halo, are discussed in \S~\ref{remarks}. Nebular conditions
derived from plasma diagnostics and the possible presence of $t^2$-effects are
discussed in \S~\ref{plasma:sec}. Then we study the observed thermal grain
emission in \S~\ref{grain:emm}. Finally we construct a photo-ionization model
of the nebula in \S~\ref{model}, which is used to study the possible presence
of dust grains inside the ionized region of the nebula. Final conclusions are
given in \S~\ref{concl}.

\section{Observations}
\label{obs:sec}

\subsection{Unpublished Data}

In the analysis of NGC~6445 several spectra were used, covering a wide range
of wavelengths. A log of all the observations that were not previously
published can be found in Table~\ref{obs:log}. The {\it International
Ultraviolet Explorer} ({\it IUE}) spectra were taken from the INES catalog in
the {\it IUE} final archive. No further processing was done on these data. For
further reference the SWP17030 spectrum will be called IUE1 and the SWP38267
spectrum IUE2. The LWP17434 spectrum does not contain any detected lines and
will not be discussed further.

\placetable{obs:log}

The optical observations were done with the Boller \& Chivens spectrograph at
the 1.52-m ESO telescope on La Silla. CCD \#24, which is a Ford Aerospace
2048L, and grating \#25 were used. At a grating angle of 8\arcdeg\ this gave
us a wavelength coverage from 391.4~nm to 967.8~nm, or 0.283~nm per pixel and
a resolution of about 0.7~nm. We were not affected by the second order
spectrum with this setting. A special technique was used to obtain
large-aperture exposures. For this the slit of the spectrograph was oriented
North-South and the tracking rate of the telescope in the East-West direction
was offset by 0.05~arcsec/s. This caused the slit to gradually drift over the
nebula, hence causing the spectrograph to integrate photons from the entire
nebula. The exposure time listed in Table~\ref{obs:log} is the effective
exposure time, i.e.\ the slit width divided by the differential tracking rate
times the number of exposures. This technique allows the optical spectrum to
be directly comparable to other large-aperture data.

The reduction was done using standard procedures in IRAF. First the spectra
were bias subtracted and flat-fielded with a normalized dome flat. The spectra
were corrected for slit illumination. The images were geometrically corrected
to rectify the spatial axis before sky subtraction. The dispersion axis is
parallel to the chip's axis well within 2 pixels along the slit. This was good
enough for our purposes as the spatial extent for the untraceable PN was much
larger than this. The standard stars were traced when extracted. The PN
spectra were extracted with a fixed aperture based on the emission line with
the largest spatial extent. The spectra were corrected for airmass and
sensitivity with the response curve based on the standard star LTT~7379.
Finally the three spectra were averaged with min-max rejection.

\placefigure{sws:spec}

The infrared data were all taken with the {\it Infrared Space Observatory}
({\it ISO}), using the Short Wavelength Spectrometer (SWS) and the Long
Wavelength Spectrometer (LWS). The data comprise two full grating scans (using
the observing templates SWS01 with scan rate 2 and LWS01 with sampling
interval 1) and a set of deeper exposures of selected individual lines (using
the SWS02 template). A detailed description of the observing templates can be
found in de Graauw et al.\markcite{dg96} (1996) for SWS and Clegg et
al.\markcite{cl96} (1996) for LWS. The SWS01 and LWS01 data were reduced from
the Auto Analysis Result produced with the {\it ISO} pipeline, version 6.0.
The reduction was done using ISAP v1.5 and LIA v7.1, and consisted of the
removal of cosmic ray hits and other artifacts and the subsequent averaging
and smoothing of the data in each band. The SWS02 data were reduced from
Standard Processed Data produced with the {\it ISO} pipeline, version 6.0. The
SWS Interactive-Analysis System was used to remove spikes attributed to
ionized-particle impacts and to align the continua.

\begin{figure}
\mbox{\epsfxsize=0.5\textwidth\epsfbox[58 314 500 688]{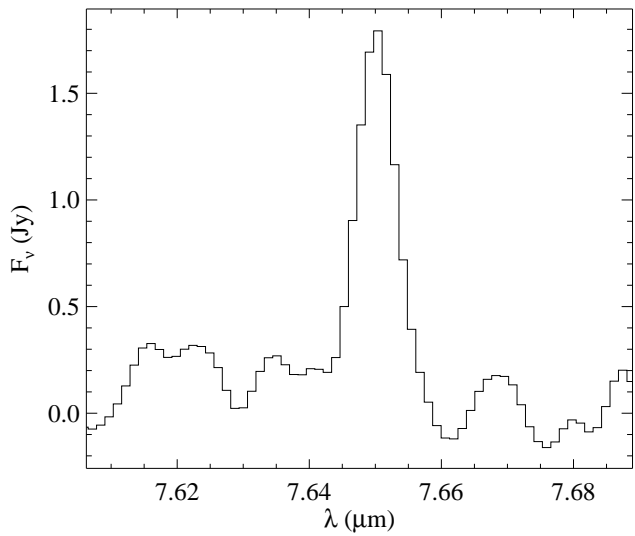}}
\figcaption[f3.ps]
{The detection of [\ion{Ne}{6}] in NGC~6445. \label{plot:ne6}}
\end{figure}

\placefigure{plot:ne6}

Almost all of the emission lines identified in the SWS and LWS spectra have
never been observed before and are a valuable extension to the UV and optical
spectra of this nebula. This is especially the case since various lines of
highly ionized atoms have been detected, which are impossible to observe from
the ground. These lines allow a more reliable determination of the stellar
temperature than was hitherto possible. To emphasize this point we show the
SWS02 detection of the [\ion{Ne}{6}] 7.65~\micron\ line in
Figure~\ref{plot:ne6} which is now the highest excitation line ever observed
in the spectrum of NGC~6445.

\subsection{Literature Data}

The optical spectrum was supplemented with data in the range from 372.6~nm to
397.0~nm taken from Aller et al.\markcite{al73} (1973). These data were
obtained at three different positions in the nebula and subsequently averaged,
so that they are a fair approximation for the integrated flux. The
[\ion{Ne}{5}] 342.6~nm flux was taken from Rowlands et al.\markcite{ro93}
(1993). These observations used a 48\arcsec\ aperture, which should be large
enough to encompass nearly all of the [\ion{Ne}{5}] flux. The adopted fluxes
are listed in Table~\ref{flux:tab}.

An H$\alpha$+[\ion{N}{2}] image of NGC~6445 was obtained by Schwarz, Corradi
\& Melnick\markcite{sc92} (1992) and is shown in Figure~\ref{aperture}. From
this image it is clear that the ionized region of NGC~6445 comprises not only
the high surface brightness region that is readily observable, but also low
surface brightness regions that extend much farther. The entire ionized region
measures at least 2\arcmin$\times$3\arcmin.

The nebula has been observed at radio frequencies on several occasions, both
using single-dish telescopes and interferometers. There is a significant
difference between the fluxes obtained with the two types of telescopes. This
is most likely due to the presence of extended low surface brightness radio
emission which was included in the beam of single dish telescopes but must
have been missed in the interferometer observations. The discrepancy between
the fluxes cannot be attributed to optical depth effects since modeling of the
nebula shows that optical depth effects only set in below 1~GHz. Hence it is
safe to assume that the radio emission is still optically thin, even at the
lowest frequencies observed with the VLA and Parkes. For the total radio flux
emitted by the nebula (including low surface brightness regions) we adopt the
14.7~GHz radio flux of 322~mJy obtained with the Parkes 64-m radio telescope
(Milne \& Aller\markcite{ma82} 1982).

The distance of NGC~6445 is not accurately known. Sabbadin\markcite{sa86}
(1986a) lists two extinction distances of 2.0~kpc and 2.5~kpc. Various other
determinations have been published in the literature which usually are lower.
We will adopt the statistical distance based on an empirical relation between
the surface brightness temperature and the radius of the nebula discussed by
Van de Steene \& Zijlstra\markcite{st95} (1995). In order to obtain a distance
with this method, we will need an angular diameter and the radio flux
contained within that radius. The Parkes data are not suitable for this
purpose, since it contains emission from low surface brightness regions and
the angular diameter of that region is not well defined. As an alternative we
will use the VLA 1.46~GHz data presented in Phillips \& Mampaso\markcite{pm88}
(1988). They obtained a radio flux of 243~mJy and an angular extent of the
high surface brightness region of 30\farcs7$\times$40\farcs2. We will adopt
the resulting geometric mean of 35\farcs1 as the angular diameter. It is
consistent to use these data and the resulting value for the distance is
1.5~kpc. This result is not very sensitive to the adopted radio flux density
$S_\nu$ since it scales as $S_\nu^{-0.35}$. Using the Parkes flux instead of
the VLA flux would yield a result that is only 15\% lower.

Using the data presented above we find that the dynamical age $t_{\rm dyn}$ of
the nebula is $t_{\rm dyn} = \Theta_{\rm r} D / v_{\rm exp}$ = 3300~yr. Here
$\Theta_{\rm r}$ stands for the angular radius of the nebula, $D$ for the
distance and $v_{\rm exp}$ for the expansion velocity.

\section{Aperture Corrections}
\label{aper:corr}

Although all of the spectral data we present have been taken with large
apertures, none of them encompass all of the nebula. This is especially the
case for the {\it IUE} and SWS observations where significant parts of the
high surface brightness regions have been missed. Therefore aperture
corrections will be necessary to use these data for modeling the nebula. The
apertures are over-plotted on an image of NGC~6445 in Figure~\ref{aperture}.

The {\it IUE}-SWP large aperture measures 9\farcs07$\times$21\farcs65 (IUE1 \&
2). SWS Data in the range between 2.38~\micron\ -- 12.0~\micron\ were taken
with an aperture of 14\arcsec$\times$20\arcsec\ (SWS-A), between 12.0~\micron\
-- 27.6~\micron\ with an aperture of 14\arcsec$\times$27\arcsec\ (SWS-B), and
between 29.5~\micron\ -- 45.2~\micron\ with an aperture of
20\arcsec$\times$33\arcsec\ (SWS-C). Data in the range between 27.6~\micron\
-- 29.5~\micron\ were taken with a 20\arcsec$\times$27\arcsec\ aperture.
However, since there are no observed lines in this range, this aperture will
be ignored in the subsequent discussion. To obtain aperture correction factors
for the {\it IUE} and SWS data, the total flux was integrated inside each of
the aperture boxes and compared to the total flux emitted by the nebula. For
this we used the H$\alpha$+[\ion{N}{2}] image of Schwarz et al.\markcite{sc92}
(1992). We obtained the following aperture correction factors which we will
adopt in the remainder of the paper. For IUE1: 15.43, for IUE2: 9.98, for
SWS-A: 5.27, for SWS-B: 4.65 and for SWS-C: 3.45.

The correction factors for the {\it IUE} data are large since this instrument
measured only a small region in the center of the nebula, missing most of the
highest surface brightness regions. This correction could have been
overestimated due to any of the following three effects. First, errors of a
few arcsecond in the satellite pointing could cause more flux to enter the
aperture. Second, light from the brightest regions just outside the aperture
could have been scattered into the line of sight by the intervening ISM.
Third, the lines {\it IUE} detected come from highly ionized species which
could be more centrally concentrated than the \ion{H}{1}+[\ion{N}{2}]
emission. The {\it IUE} fluxes should therefore be considered more uncertain
than any of the other data.

The fact that the SWS apertures are centered on a surface brightness peak
assures that the first two effects will have less influence on the correction
factors for these data. Ionization stratification could however still be
important. To study this further, we inspected the [\ion{O}{3}] image taken by
Schwarz et al.\markcite{sc92} (1992). The correction factors for the SWS data
derived from this image agree with the ones listed above at the 10\% level.
This can be explained by the fact that the positioning of the apertures (which
include both ionized and neutral material) favors a small effect on the
correction factors. We therefore conclude that the aperture corrections for
the SWS data are fairly reliable.

The apertures for the optical and {\it ISO} LWS observations are much bigger
than the {\it IUE} and {\it ISO} SWS apertures, and cover all of the high
surface brightness regions of the nebula. We will not apply any aperture
corrections to these data.

\section{Interstellar Reddening}
\label{reddening}

The values for the interstellar extinction towards NGC~6445 quoted in the
literature vary considerably and should therefore be considered uncertain. In
view of this fact and also because we have spectra covering a wide range of
wavelengths, a careful analysis of the data is warranted in order to obtain an
accurate extinction curve. Such an analysis is necessarily an iterative
process: in order to obtain an extinction curve we need estimates for the
electron temperature and density, which in turn can only be determined
accurately after an extinction correction has been applied. Luckily the
dependence on these quantities is only moderate to weak and simple estimates
are sufficient. In the following paragraphs we adopt values for the electron
temperature and density which do not agree precisely with each other or with
the final values which will be derived further on in the paper. We are however
convinced that the assumed values are sufficiently accurate to ensure that the
errors introduced by these assumptions are far less than any other source of
error.

We will start the analysis by comparing the radio and H$\beta$ fluxes. Since
we estimate that the optical aperture encompasses roughly 90\% of the total
flux emitted by the nebula, it is appropriate to use the Parkes 14.7~GHz flux
of 322$\pm$18~mJy (Milne \& Aller\markcite{ma82} 1982) for this comparison.
Substituting this value in Equation \mbox{IV-26} of Pottasch\markcite{po84}
(1984), and adopting $Y = 1.23\pm0.05$ and an average electron temperature of
$T_{\rm e} = 13.8\pm1.3$~kK (obtained from a preliminary analysis of the
[\ion{O}{3}] lines), we were able to obtain an estimate for the dereddened
(intrinsic) H$\beta$ flux: $\log[F(H\beta)$/(W\,m$^{-2})] = -13.06\pm0.04$.
When we compare this with the observed H$\beta$ flux of
$\log[F(H\beta)$/(W\,m$^{-2})] = -14.12\pm0.04$, we find that $c({\rm H}\beta)
= 1.06\pm0.06$~dex, or alternatively $A(486.1~\nm) = 2.64$~mag.

\begin{figure}
\mbox{\epsfxsize=0.7\textwidth\epsfbox[37 314 535 688]{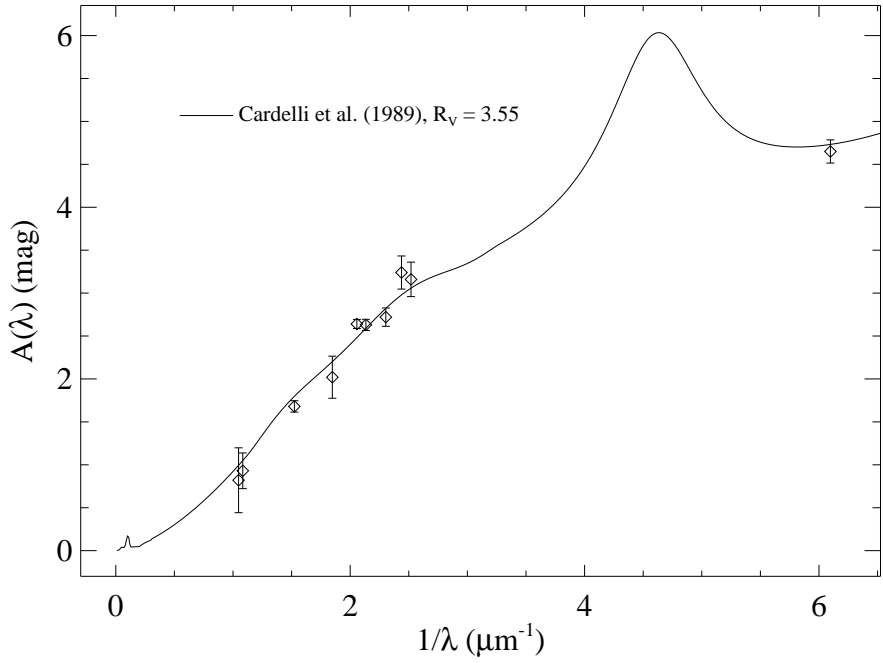}}
\figcaption[f4.ps]
{The extinction law for NGC~6445. \label{plot:abs}}
\end{figure}

We will now determine the shape of the extinction curve for our spectra. The
relative line strengths in the recombination spectrum of \ion{H}{1} and
\ion{He}{2} depend very little on electron temperature and density and can be
determined using Case~B theory. This allows us to determine the absolute
extinction at various other wavelengths using the value for $A(486.1~\nm)$. We
adopt the intensity ratios calculated by Storey \& Hummer\markcite{sh95}
(1995), assuming an electron density of $10^3$~cm$^{-3}$ and an electron
temperature of 12.5~kK and 15~kK for the \ion{H}{1} and \ion{He}{2} emitting
regions respectively. These assumptions introduce an error of approximately
1\% or less in the Case~B line ratios and are therefore well justified. The
results are shown in Table~\ref{obs:abs}. The Case~B ratios are normalized to
\ion{H}{1} 486.1~nm and \ion{He}{2} 468.6~nm for the hydrogen and helium
spectra respectively. In order to obtain a value for the extinction at
468.6~nm, a fit was made to the observed extinctions for the hydrogen lines
only, using a standard ISM extinction law (Cardelli, Clayton \&
Mathis\markcite{ca89} 1989), and adopting a ratio of the total to specific
extinction $R_V \equiv A(V)/E(B-V) = 3.3$. This allowed us to obtain an
absolute flux of the \ion{He}{2} 468.6~nm line and include the other
\ion{He}{2} lines in the analysis. The results are shown in
Table~\ref{obs:abs} and Figure~\ref{plot:abs}.

\placetable{obs:abs}

\placefigure{plot:abs}

To obtain an extinction curve, we made a least-squares fit of the law given by
Cardelli et~al.\markcite{ca89} (1989) to these data, using the ratio of total
to selective extinction, $R_V$, and the selective extinction, $E(B-V)$, as
free parameters. The result was $R_V = 3.55$ and $E(B-V) = 0.61$, which is
equivalent to $c({\rm H}\beta) = 0.99$~dex. We have composed an extinction law
using the results of Cardelli et al.\markcite{ca89} (1989) with $R_V = 3.55$
for wavelengths between 0.1~\micron\ and 3.33~\micron, Rieke \&
Lebofsky\markcite{rl85} (1985) between 3.33~\micron\ and 13~\micron, and
Mathis\markcite{ma90} (1990) for wavelengths longward of 13~\micron. The
extinction law that is obtained this way is plotted in Figure~\ref{plot:abs}
and was used to deredden all observed fluxes.

Altering the value of $R_V$ has a strong effect on the UV part of the
extinction curve, but leaves the optical and infrared part virtually
unchanged. Hence the fact that we use an $R_V$ which deviates from the
canonical value $R_V = 3.1$ is important for the dereddening of the UV line
fluxes and assures that the Case~B line ratios for \ion{He}{2} are obeyed. On
the other hand, if we had used an extinction law with $R_V = 3.1$, this would
only have made a difference of 10\% or less for the optical and infrared line
intensities. In other words, in cases where only optical and/or infrared data
are available, the value of $R_V$ cannot be determined from these data and
assuming $R_V = 3.1$ is appropriate. But in cases where UV data is included,
$R_V$ should be treated as a free parameter to assure a proper dereddening.

The results of the dereddening can be found in Table~\ref{flux:tab} where we
list the fluxes and the dereddened intensities relative to H$\beta$ for all
the lines used in this paper. Aperture correction factors have been applied to
the dereddened intensities, but not to the observed fluxes (except for the
combined {\it IUE} spectra). The error bars quoted in Table~\ref{flux:tab}
reflect purely the uncertainty in the measured flux and are therefore a
measure for the quality of the spectrum. They do {\em not} contain
contributions for the uncertainty in the absolute flux calibration, aperture
correction or reddening correction. However, estimates for these uncertainties
where quadratically added to make the photo-ionization model discussed in
\S~\ref{model}.

\placetable{flux:tab}

\section{Remarks on the Observed Spectrum}
\label{remarks}

The observed flux for the [\ion{O}{3}] 88.2~\micron\ line is more than a
factor of 5 higher than the 3$\sigma$ upper limit of
2$\times$10$^{-14}$~W\,m$^{-2}$ that was previously determined by Dinerstein,
Lester \& Werner\markcite{di85} (1985). This large discrepancy cannot be
explained by aperture effects since both observations encompass essentially
all of the nebula. It also seems very unlikely that it is caused by
variability in the spectrum since there is no evidence that would support this
assumption. Comparison of our [\ion{O}{3}] 500.7~nm flux with the ones
published in O'Dell\markcite{od63} (1963) and Aller et al.\markcite{al73}
(1973) shows that they all are in good agreement. Our flux is only 0.06~dex
higher than the older measurements, making it very unlikely that the infrared
[\ion{O}{3}] spectrum has undergone large changes over the past 35 years. An
explanation for this conundrum is therefore not apparent to us.

The intensity $I({\rm H}\alpha) = 363$ is surprisingly large and is not
expected by Case~B theory. A possible explanation could be provided by the
fact that the overall shape of the extinction curve can differ from the mean
law for different lines of sight (Cardelli et al.\markcite{ca89} 1989). HST
observations could be used to investigate the extinction law further.

\begin{figure}
\mbox{\epsfxsize=0.6\textwidth\epsfbox[54 360 528 719]{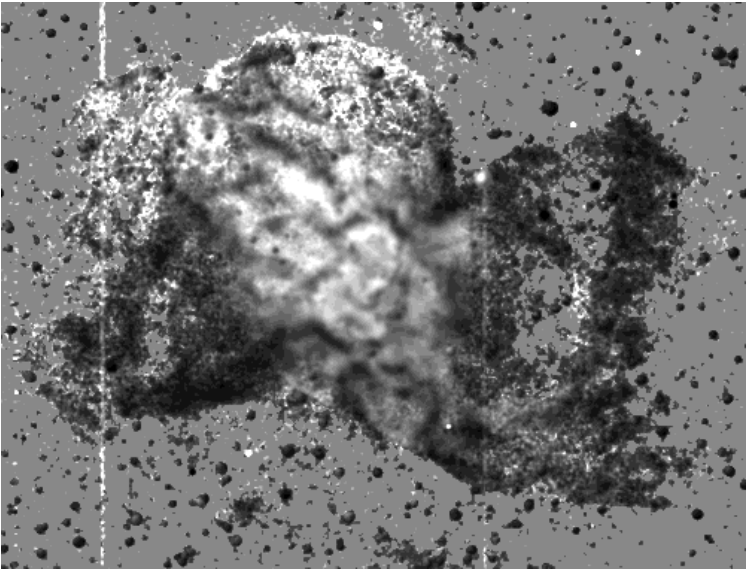}}
\figcaption[f5.ps]
{The ratio of the [\ion{O}{3}] over H$\alpha$+[\ion{N}{2}] image of
NGC~6445 (Schwarz et al.\protect\markcite{sc92} 1992).
Dark regions indicate low excitation and are interpreted as regions near
an ionization front or unilluminated regions that are currently recombining.
\label{ratio}}
\end{figure}

\placefigure{ratio}

Another explanation could be provided by the fact that the outer halo of
NGC~6445 is currently recombining (Tylenda\markcite{ty86} 1986), which would
cause the plasma to have an electron temperature well below what is expected
for an ionized region and would alter the Case~B line ratios. From our
discussion in \S~\ref{model} it will become clear that the central star is
currently evolving on the cooling track in the Hertzsprung-Russell diagram,
which implies that it has undergone a considerable drop in luminosity in
recent history. This evolution could have caused the ionized region to shrink.
Therefore it is possible that the faint surface brightness regions are
actually the afterglow of regions that did not have sufficient time to
recombine. The recombination timescale for hydrogen is approximately
$1230\,T_4^{0.83} n_2^{-1}$~yr, where $T_4$ is the electron temperature in
units of $10^4$~K and $n_2$ is the electron density in units of
$10^2$~cm$^{-3}$ (using Storey \& Hummer\markcite{sh95} 1995). This timescale
could very well be longer than the evolutionary timescale of the central star,
depending on the density in the outer halo. However, the timescale for
recombination from O$^{2+}$ to O$^+$ is only $86\,T_4^{0.52} n_2^{-1}$~yr
(using the atomic data referenced in {\sc cloudy} 90.05). This would imply
that in the outermost regions of the halo, where the gas supposedly had the
longest time to recombine, one would not expect [\ion{O}{3}] 500.7~nm emission
to be visible. Inspecting the [\ion{O}{3}] image taken by Schwarz et
al.\markcite{sc92} (1992) however reveals that such emissions are present. The
outermost regions in the North-East and South-West directions are clearly
visible in both the H$\alpha$+[\ion{N}{2}] and the [\ion{O}{3}] image,
indicating that they are still being photo-ionized. Some regions in the
South-East and North-West directions however do appear much fainter in
[\ion{O}{3}], and it is plausible that recombination is going on there. To
illustrate this point, we show in Figure~\ref{ratio} the ratio of the
[\ion{O}{3}] over the H$\alpha$+[\ion{N}{2}] image. It can be interpreted as a
measure for the level of excitation, where white indicates high excitation.
The scale and positioning of the image is identical to Figure~\ref{aperture}.
In order to increase the signal-to-noise ratio in the low surface brightness
regions, the original images were smoothed with a gaussian with a FWHM of 2
pixels before they were divided. The resulting image clearly shows the complex
structure of the nebula. The darkest regions in the image are either low
excitation regions that are close to an ionization front or unilluminated
regions where recombination is taking place. The image clearly shows that
especially the North-East part of the halo is still highly excited. Hence the
simple view where the entire halo is recombining cannot be correct. Due to the
evolutionary status of the central star it is likely however that
recombination is taking place to some degree. This is the case for the
South-East and North-West regions, but probably near every ionization front as
well. We cannot obtain an accurate estimate for the emission measure of these
regions, but it seems likely it is insufficient to have a large effect on the
integrated hydrogen spectrum, at least within the optical aperture. On the
other hand, recombination is expected to have a profound effect on the
conditions in the Photo-Dissociation Region (PDR). This is because the region
which is now the PDR, was in recent history part of the ionized region and
therefore still has a `memory' of those conditions. In particular, its energy
content will be higher than expected from equilibrium calculations, and
therefore also the electron temperature will be higher. This implies that {\sc
cloudy} may not be able to correctly predict the PDR emissions for this
nebula.

The fact that ionization fronts can be seen in virtually every direction in
Figure~\ref{ratio} is a strong support for the assumption that the nebula is
ionization bounded and that virtually no Lyman continuum photons can escape
it. That the nebula is ionization bounded is possible because the central star
is on the cooling track and emits far less ionizing photons than when it was
still on the horizontal track. The fact that ionization fronts can clearly be
seen in the outer halo proves that the halo cannot be a reflection nebula.

\section{Plasma Diagnostics and the $t^2$-effect}
\label{plasma:sec}

Accurate abundance determinations are hampered by the so-called $t^2$-problem
(Peimbert\markcite{pe67} 1967). This problem is probably one of the most
important in present day PN research and subject to much debate. The
$t^2$-parameter is a measure for the electron temperature fluctuations in the
ionized region of the nebula. Models do not predict such fluctuations, but the
limited amount of observational material gathered so far seems to contradict
this. The $t^2$-problem can be investigated by comparing the strength of
infrared lines with optical lines. The emissivities of the far-infrared
fine-structure lines have a very different dependence on electron temperature
and density as compared with the optical lines (Simpson\markcite{si75} 1975;
Watson \& Storey\markcite{ws80} 1980). The infrared line-strengths are
virtually insensitive to $T_{\rm e}$, because of the very small energy of
their upper levels, whereas the optical lines originate from higher energy
levels and have a strong dependence on $T_{\rm e}$. Fluctuations in $T_{\rm
e}$ increase the strength of emission lines and this increase will be larger
for lines originating from higher energy levels (e.g., for optical lines).
Hence the $t^2$-effect can be measured by comparing electron temperatures
determined from optical and infrared diagnostic lines.

Using the dereddened line intensities, we are able to derive values for the
electron temperature and density using various diagnostic ratios. The results
are listed in Table~\ref{plasma:diag}. The results for the optical density
diagnostics ([\ion{S}{2}] and [\ion{O}{2}]) were obtained with the package
{\sc nebular} in {\sc iraf} (Shaw \& Dufour\markcite{sd95} 1995). All other
diagnostics were obtained using our own five level model atom, with collision
strengths taken from Lennon \& Burke\markcite{lb94} (1994) for [\ion{N}{2}],
[\ion{O}{3}] and [\ion{Ne}{5}], Butler \& Zeippen\markcite{bz94} (1994) for
[\ion{Ne}{3}], and Galavis, Mendoza \& Zeippen\markcite{ga95} (1995) for
[\ion{S}{3}]. The transition probabilities were taken from Galavis, Mendoza \&
Zeippen\markcite{ga97} (1997) for [\ion{N}{2}], [\ion{O}{3}], [\ion{Ne}{3}]
and [\ion{Ne}{5}], and Mendoza \& Zeippen\markcite{mz82} (1982) for
[\ion{S}{3}].

\placetable{plasma:diag}

The results for the various temperature diagnostics show quite a large spread,
which can in part be explained by temperature stratification in the nebula.
However, one also notices that the infrared temperature diagnostics seem to
give systematically lower results than the optical diagnostics. This could
point to a $t^2$-effect, although the evidence is not convincing. The
temperatures we derived indicate the following values: $t^2$[\ion{O}{3}] =
0.065$\pm$0.053 and $t^2$[\ion{S}{3}] = 0.168$\pm$0.148. Hence the error
margins are too large to validate the presence of a $t^2$-effect.

Other explanations for the difference between the optical and infrared
diagnostics are also plausible. First of all, in order to calculate the
temperature from the infrared diagnostics, the optical and {\it ISO} spectra
need to be compared. Hence systematic effects may be introduced by errors in
the absolute calibration of either of these spectra and/or by errors in the
aperture corrections for the SWS data. The uncertainty introduced by the
absolute calibrations has been estimated at 20\%, and the uncertainty
introduced by the aperture corrections at 15\%. These numbers were used to
calculate the error margins for the electron temperatures given above, but
they may have been underestimated.

Another possible explanation is inaccuracies in the atomic data, in particular
the collisional cross sections. Questions concerning the [\ion{Ne}{5}] data
calculated by Lennon \& Burke\markcite{lb94} (1994) have been raised by Oliva,
Pasquali \& Reconditi\markcite{ol96} (1996). However, their argument can be
refuted as it is based on an inaccurate flux for the [\ion{Ne}{5}]
14.3~\micron\ line, as is shown by van Hoof et al.\markcite{vh97b} (1997b).
Subsequent work (van Hoof et al.\markcite{vh99} 1999) has shown that, at least
for [\ion{Ne}{5}], deviations up to a maximum of 30\% in the data for the
transitions in the ground term could be indicated by observations of PNe. For
the data in this paper, the difference between the optical and infrared
diagnostics can be eliminated by adjusting the collision strengths of the
transitions in the ground term at the 30\% level or less. This argument
indicates that it is at least conceivable that the differences between the
optical and infrared diagnostics are caused by inaccuracies in the collisional
cross sections.

The various density diagnostics are in excellent agreement, which suggests
that the assumption of constant density is a good approximation for modeling
the ionized region of this nebula. The density we find is fairly low for
planetary nebulae, which indicates that NGC~6445 is well evolved.

\section{Thermal Grain Emission}
\label{grain:emm}

The fact that hardly any grain emission is detected in the SWS01 spectrum
raises the question whether the inner regions of NGC~6445 could be free of
grains, either due to grain destruction, or to dust-gas separation. The LWS
spectrum covers a larger part of the nebula, and grain emission is observed in
this spectrum, which indicates that at least part of the grains survived. LWS
will have detected the emission from the grains closest to the central star,
but still may have missed the emission from the cooler grains farther out in
the nebula. Before we can study this spectrum, it needs to be corrected for
the contribution from the ISM background.

\subsection{Correcting the On-Source Spectrum}

\begin{figure}
\mbox{\epsfxsize=0.7\textwidth\epsfbox[36 314 535 688]{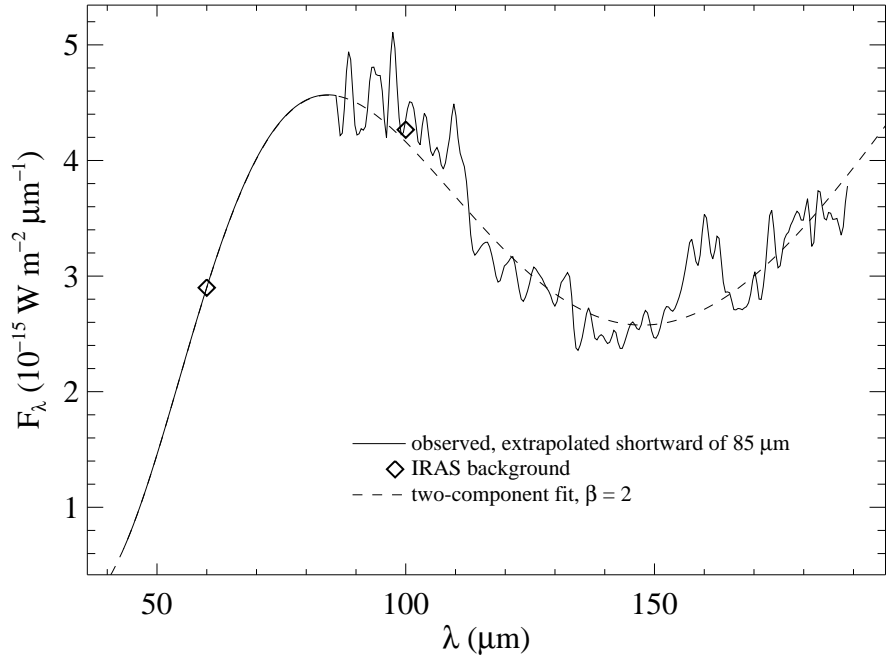}}
\figcaption[f6.ps]
{The off-source spectrum near NGC~6445. \label{off:source}}
\end{figure}

We took an off-source LWS spectrum 3\arcmin\ north of the nebula at a position
angle of roughly $-50^\circ$ with respect to the polar axis of the nebula
(using the standard convention where positive angles indicate a
counter-clockwise direction). The approximate position of the polar axis is
indicated in Figure~\ref{aperture}. The off-source spectrum was intended to
measure the background continuum and line emission from the ISM. The
off-source spectrum is shown in Figure~\ref{off:source}. It shows two clearly
distinct emission components: one very cool component which only becomes
detectable at wavelengths longward of 140~\micron, and a much warmer component
which is visible at shorter wavelengths. The latter is observable as a smooth
background in HIRES images made from IRAS 60~\micron\ and 100~\micron\ data.
Measurements of this background are shown with a diamond in
Figure~\ref{off:source}. The flux densities were calculated from the measured
surface brightness assuming a circular aperture with a diameter of 84\arcsec\
for LWS and applying the appropriate AC-DC and color corrections. The
excellent agreement of the 100~\micron\ point clearly indicates that the warm
component of the background emission is associated with cirrus. No significant
gradient of the cirrus emission is seen in the vicinity of NGC~6445. The very
cold component is not detected in the 100~\micron\ HIRES image since its flux
is too low compared to the cirrus emission. The cold component is clearly
extended (it is seen in both the on-source and off-source spectrum). Estimates
for the dust temperature of both components are derived below. The quality of
the off-source spectrum was poor at wavelengths shortward of 85~\micron\ and
therefore an extrapolation was made using the fit discussed below. Hence the
extrapolation is primarily based on the background flux measured in the
60~\micron\ HIRES image. The ISM contribution is very small compared to the
on-source flux at 45~\micron\ (less than 2\%). For longer wavelengths the ISM
contribution becomes increasingly important: at 85~\micron\ the ISM
contributes 25\% to the total on-source flux, at 145~\micron\ 50\%, and at
190~\micron\ 90\%.

\placefigure{off:source}

\begin{figure}
\mbox{\epsfxsize=0.7\textwidth\epsfbox[22 314 532 688]{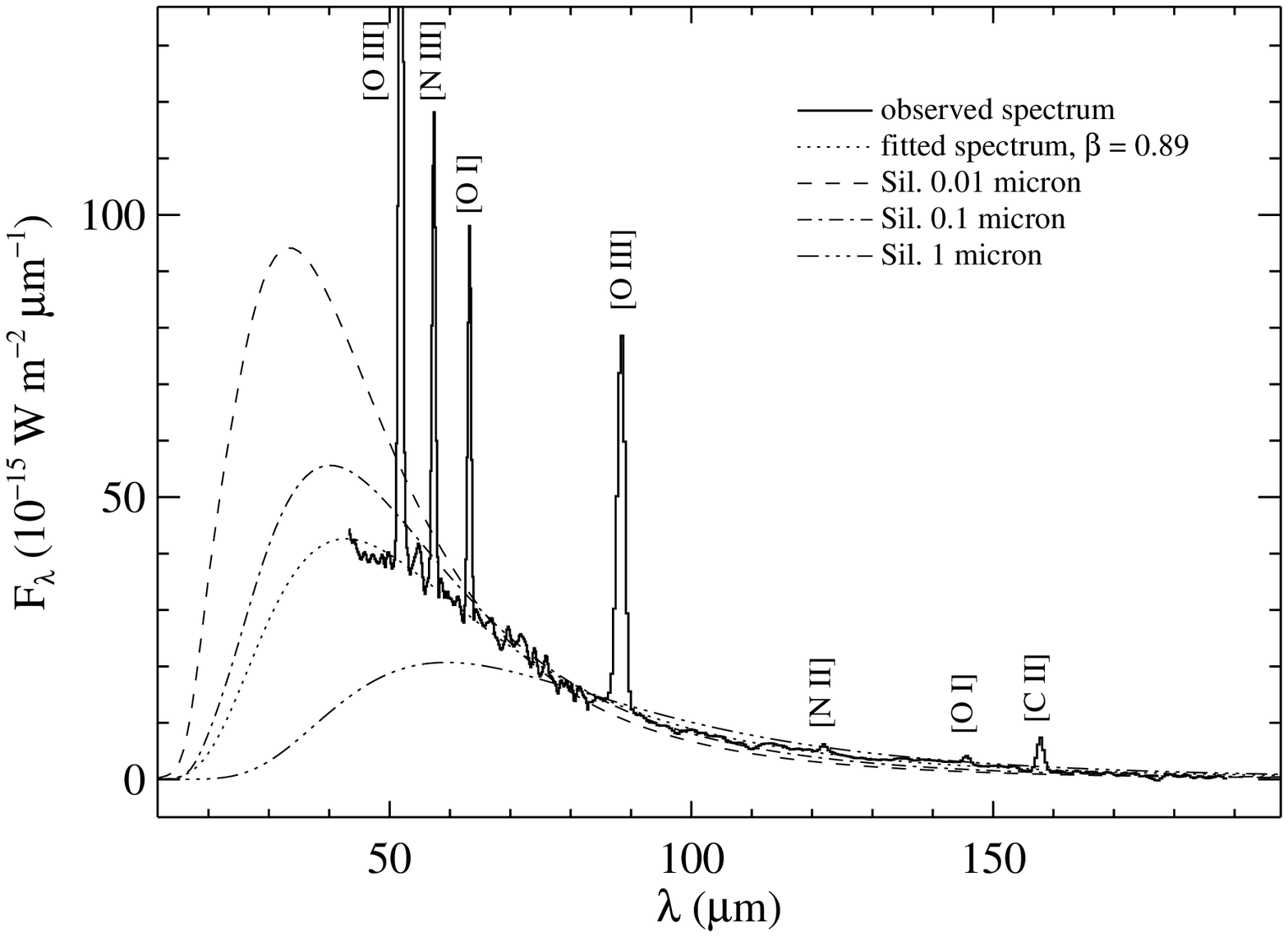}}
\figcaption[f7.ps]
{The far-infrared spectrum of NGC~6445. \label{plot:farir}}
\end{figure}

We need to correct the on-source spectrum for the contribution by the ISM
background since this emission would lead to a serious underestimation of the
nebular grain temperature. This correction is easily done assuming that the
contribution by the ISM is the same in the on-source and off-source spectrum.
This assumption is justified, both for the warm and cold component of the ISM
emission. From the HIRES images we know that the cirrus emission shows no
gradient in the vicinity of the source. The very cold ISM component is the
dominant feature at long wavelengths in the on-source spectrum and this
feature is very nicely removed after subtraction. The remaining question is if
grain emission from the outermost regions of the nebula itself could have
contributed to the off-source spectrum. This argument leads to estimates of
hundreds or even thousands of solar masses for the total nebular mass. This is
clearly impossible unless we have seriously overestimated the distance to the
nebula. This argument seems very contrived and we therefore dismiss it. The
on-source spectrum was corrected for the contribution of the ISM continuum
using the entire spectrum shown in Figure~\ref{off:source}. The corrected
spectrum is shown in Figure~\ref{plot:farir}. In the off-source spectrum two
emission lines were detected, [\ion{O}{1}] 63~\micron\ and [\ion{C}{2}]
157~\micron. The fluxes of these lines were subtracted from the on-source flux
to constrain our photo-ionization model of the nebula.

\placefigure{plot:farir}

\subsection{Temperature Determination}

To determine the average grain temperature of the ISM background emission and
the nebular grain emission, we will use an emission law of the form $F_\nu = c
\, \nu^\beta B_\nu(T_{\rm dust})$ where $c$, $\beta$ and $T_{\rm dust}$ are
free parameters. Such a simple power law dependence of the absorption cross
section is assumed for astronomical (i.e., amorphous) silicates beyond the
18~\micron\ feature and for graphite beyond 35~\micron\ (e.g., Draine \&
Lee\markcite{dl84} 1984). However, recent {\it ISO} observations have shown
that crystalline silicates show a rich spectrum of emission features, both in
the SWS and LWS wavelength range (e.g., Waters et al.\markcite{wa96} 1996),
and a power law approximation is not valid for such grains. In the on-source
spectrum we do not detect any crystalline silicate features in the LWS range,
but we cannot rule out the presence of such features below the detection limit
in the SWS observations. In the absence of better data, we will assume that
the power law approximation is valid in the LWS range. This assumption is
commonly made for fitting diffuse ISM emissions, often using $\beta = 2$.

\subsubsection{The ISM Background}

Before we discuss the grain emission from the nebula itself, we discuss the
fits to the ISM background emission. We derived estimates for the temperature
of the warm and very cold component by using a two-component model. Each
component has the same analytic dependence given above and we assumed $\beta =
2$ for both components. To constrain the fit we used the 60~\micron\ point
from the HIRES image and the LWS off-source spectrum longward of 85~\micron.
The optimal fit is shown in Figure~\ref{off:source} and Table~\ref{farir:tab}.
We find that the warm cirrus component has a temperature of 24~K, which makes
it possibly the warmest cirrus cloud known. Lagache et al.\markcite{la98}
(1998) show various measurements in their Figure~2, the highest being 23~K in
the Orion complex. The galactic coordinates for our measurement are $l_{\rm
II} = 8.119$, $b_{\rm II} = +3.930$. Comparison with Figure~1 of Lagache et
al.\markcite{la98} (1998) shows that they are in the middle of a
high-temperature region. Since their data is smoothed over a 7$^\circ$ beam,
small clouds with higher temperatures could have been averaged out. Taken at
face value, our detection of a very cold component of 7~K is in agreement the
results of Wright et al.\markcite{wr91} (1991) and Reach et al.\markcite{re95}
(1995), who derived the widespread presence of such a component based on FIRAS
data taken by {\it COBE}. Lagache et al.\markcite{la98} (1998) subsequently
dismissed these claims, upon a re-interpretation of DIRBE and FIRAS data.
Since our observation encompasses only a small region of the sky, it is not
clear how extended the 7~K component is and whether it could have contributed
significantly to the FIRAS spectrum. Note also that our line of sight is in
the galactic plane. Given this fact, the very cold emission component may come
from a starless core. Ward-Thompson et al.\markcite{wt94} (1994) have observed
two such cores with temperatures of 8~K. Bernard et al.\markcite{be92} (1992)
predict that for a core to reach a temperature as low as 7~K, the optical
depth should at least be $A_V = 100$~mag. So for NGC~6445 to be observable,
the core should reside behind the PN, giving a minimum distance of 1.5~kpc.
Because we see the very cold component in both the on-source and off-source
spectrum, the angular extent must be at least 5\arcmin. This leads to a lower
limit of 2~pc for the diameter of the core. This is considerably larger than
the cores observed by Ward-Thompson et al.\markcite{wt94} (1994). It is not
clear whether cores this large can sustain a temperature of 7~K (see e.g.\ the
discussion in Laureijs\markcite{la99} 1999, and references therein). The
ISOPHOT Serendipity Survey (ISOSS) also finds very cold cores with
temperatures of approximately 13~K and a radial extent consistent with our
observations (Hotzel et al.\markcite{ho99} 1999). It should be noted that
ISOSS can only detect cores with temperatures in the range 10~K $< T_{\rm
dust} <$ 70~K (T\'oth et al.\markcite{to99} 1999). Hence the non-detection of
cores with temperatures below 10~K in ISOSS is expected. Further observations
are needed to clarify the nature of the very cold component.

\placetable{farir:tab}

\subsubsection{The Nebular Grains}

The LWS data for NGC~6445 show evidence for the presence of dust which is
cooler than observed in most other planetary nebulae. We will investigate this
emission first, before we try to determine where the emitting grains reside.
The resulting best fit gave $\beta = 0.89$ and is shown in
Table~\ref{farir:tab} and Figure~\ref{plot:farir}. Fits with $\beta$ fixed to
other values are also shown for comparison in Table~\ref{farir:tab}. The total
flux emitted by the grains, F(IR) = 2.38$\times$10$^{-12}$~W\,m$^{-2}$, is
nearly a factor of two lower than was previously reported by Lenzuni, Natta \&
Panagia\markcite{le89} (1989) and Zhang \& Kwok\markcite{zk93} (1993). This
discrepancy can be attributed to the fact that both papers based their
determination on the IRAS fluxes in all four bands. However, the IRAS
12~\micron\ and 25~\micron\ detections of NGC~6445 are strongly dominated by
line emission, which leads to an overestimation of the total infrared flux.

The temperature of 57~K for the nebular dust is considerably lower than what
is typically observed in planetary nebulae. Dust grains with such typical
temperatures (roughly between 90~K and 150~K) cannot be present in NGC~6445,
since they would be readily observable in the SWS spectrum. The ratio of the
thermal grain emission to the H\,$\beta$ flux of 27.0 is also very low. The
presence of strong neutral lines in the spectrum indicates that the nebula is
ionization bounded (see also \S~\ref{remarks}). This argument is further
strengthened by the fact that most determinations of the hydrogen and helium
Zanstra temperatures for the central star are in good agreement (see
Table~\ref{cst:tab}). This is a good indication that the hydrogen Lyman
continuum is optically thick. We therefore assume that no significant fraction
of the Lyman continuum photons escape the nebula. Our photo-ionization model
(see \S~\ref{model}) predicts $F$(Ly\,$\alpha$)/$F$(H\,$\beta$) = 26.6, which
implies that all of the grain heating could be accounted for by Ly\,$\alpha$
photons. Apparently Lyman continuum photons contribute little to grain heating
and this suggests either that the ionized region is free of grains, or that
the ionization parameter is low, resulting in a low optical depth of the
grains in the Lyman continuum (Bottorff et al.\markcite{bo98} 1998). To
investigate this further we will first construct a photo-ionization model of
the ionized region of NGC~6445.

\section{Photo-ionization Modeling}
\label{model}

We constructed a photo-ionization model of the ionized region using a modified
version of {\sc cloudy} 90.05 (Ferland et al.\markcite{fe98} 1998) and
following the method described in van Hoof \& Van de Steene\markcite{vv99}
(1999) with the following alterations. Instead of the blackbody assumption we
used model atmospheres calculated by Rauch\markcite{ra97} (1997). Furthermore,
we used a constant hydrogen density throughout the ionized and neutral regions
of the nebula. In view of the complex structure of the nebula we did not
constrain the model with an angular diameter, but alternatively fixed the
density at the value determined from plasma diagnostics. We also used the
electron temperatures determined from various plasma diagnostics as additional
constraints on the model. We assumed no dust grains to be present in the
ionized region and did not constrain the model with infrared continuum fluxes.
We omitted the infrared fine-structure lines of [\ion{O}{1}] from the modeling
input since preliminary models showed a large discrepancy between the modeled
and observed flux of these lines. The observed fluxes are approximately a
factor of 3 to 5 stronger than the model fluxes and we feared that such a
large discrepancy would upset the abundance determination. A possible
explanation for this discrepancy could be provided by the fact that parts of
the nebula are recombining, which would have a strong influence on the PDR
physics. {\sc Cloudy} is not capable of modeling such non-equilibrium
conditions, and an underestimation of the flux of the [\ion{O}{1}]
fine-structure lines is expected because {\sc cloudy} will underestimate the
electron temperature in the recombining regions. See also the discussion in
\S~\ref{remarks}.

An alternative explanation could be the fact that the collision strengths for
the [\ion{O}{1}] fine-structure lines are poorly determined because they are
difficult to calculate (Bhatia \& Kastner\markcite{bk95} 1995). For instance,
cross sections for O$^0$+e collisions have recently been revised downwards by
a factor of 5 to 50 (Bell et al.\markcite{be98} 1998), based on calculations
only valid below 3000~K. Cross sections for O$^0$+H and O$^0$+He collisions
are only available for temperatures below 1000~K (respectively Launay \&
Roueff\markcite{lr77} 1977 and Monteiro \& Flower\markcite{mf87} 1987). Since
the [\ion{O}{1}] fine-structure lines are mainly formed in regions with
electron temperatures between 4000~K and 10\,000~K, significant extrapolations
of these (possibly uncertain) data are needed. This makes modeling of the
[\ion{O}{1}] fine-structure lines very difficult and may very well explain the
discrepancy between the observed and calculated values.

The parameters for the photo-ionization model we derived are listed in
Table~\ref{mod:par}.

\placetable{mod:par}

\subsection{The Central Star}

The temperature of the central star of NGC~6445 has been determined many
times, mainly using the Zanstra method. An overview of all recently published
determinations can be found in Table~\ref{cst:tab}. Most of these are based on
a blackbody approximation. The average of all determinations using the Zanstra
method, or a variation thereof, is 188$\pm$5~kK. This is in excellent
agreement with the value that has been derived in this work, based on
atmosphere models by Rauch\markcite{ra97} (1997). This indicates that in the
case of NGC~6445 the Zanstra method works well. This can be understood given
the fact that the nebula is ionization bounded and optical depth effects due
to dust grains are negligible. Hence the basic assumptions of the Zanstra
method are valid in this case and the main uncertainties in applying the
method will stem from observational error and to a much lesser amount from
assumptions concerning the shape of the central star spectrum (i.e., the
blackbody assumption).

\placetable{cst:tab}

In order to check our determination of the luminosity of the central star, we
also made predictions for the magnitude of the central star in various
photometric bands. A comparison of the results with observed magnitudes is
shown in Table~\ref{mstar:tab}. The largest discrepancy is with the $V$
magnitude measured by Jacoby \& Kaler\markcite{jk89} (1989). Their flux may be
too low due to the possible presence of clouds during the observations.
Gathier \& Pottasch\markcite{gp88} (1988) used a narrow-band filter which is
free of emission lines. This technique is intrinsically more accurate than
using a broad-band filter like Johnson $V$. Since their measurement is in
excellent agreement with the predictions, this confirms our determination of
the central star luminosity with photo-ionization modeling.

\placetable{mstar:tab}

When one compares the stellar temperature and luminosity with the theoretical
calculations of Bl\"ocker\markcite{bl95a}\markcite{bl95b} (1995a,b), one finds
that they indicate a core mass between 0.625~\zm\ and 0.696~\zm, or
alternatively a main-sequence mass between 3~\zm\ and 4~\zm. The evolutionary
age of the nebula predicted by the Bl\"ocker tracks is approximately 3500~yr
and 900~yr for the 0.625~\zm\ and 0.696~\zm\ tracks respectively. This is in
excellent agreement with the derived dynamical age of 3300~yr.

\subsection{Dust Grains in the Ionized Region}

Our model shows the carbon and oxygen abundances to be about equal. In view of
the fact that the carbon abundance may be overestimated due to the
uncertainties in the {\it IUE} data, it seems more likely that the nebula is
oxygen-rich. However, we take no final standpoint on this issue and leave both
possibilities open. It is a well-known fact that in an oxygen-rich environment
silicate grains can be formed. These grains predominantly contain oxides of
magnesium, silicon and iron, but also of other element such as calcium,
titanium, chromium, nickel, etc. It is a lesser known fact that these elements
can also be depleted in a carbon-rich environment. See for instance Beintema
et al.\markcite{be96} (1996) where it is shown that sodium, magnesium,
aluminum, silicon, calcium and iron are depleted in the carbon-rich nebula
NGC~7027. It is unclear what types of grains these elements are depleted into.
They may either be carbides (e.g., SiC has been observed in various objects),
or they may be locked up as impurities in more abundant amorphous carbon or
graphite grains. Which of these grain types are present in NGC~6445 is
unclear, so we will simply refer to them as silicon-bearing grains.

First we will investigate whether grain destruction could have taken place in
the ionized region of NGC~6445. When dust grains are destroyed, the
constituent material is returned to the gas phase, making the composition
essentially solar again. This has a profound effect on the spectrum being
emitted by the plasma. Many strong lines of elements like magnesium, aluminum,
calcium, vanadium, chromium, etc.\ should be easily detectable in UV or
optical spectra (Kingdon, Ferland \& Feibelman\markcite{ki95} 1995; Kingdon \&
Ferland\markcite{ki97} 1997). The fact that none of these lines are detected
in the spectrum of NGC~6445 already indicates that no substantial amount of
grain destruction (at least for the silicon-bearing grains) can have taken
place in the ionized region. This is confirmed by our photo-ionization
modeling which clearly indicates that elements like magnesium, silicon and
calcium are depleted in the ionized region. The derived abundance for
magnesium is marked uncertain because it is based on only one emission line
belonging to a subordinate ionization stage, which implies that a substantial
correction is needed for unobserved ionization stages. By itself the magnesium
under-abundance is not convincing evidence for depletion. It is however
supported by the under-abundance for silicon which is better determined. It is
also based on only one emission line, [\ion{Si}{2}] 34.8~\micron, but since
Si$^+$ is the dominant ionization stage, only a small correction for
unobserved ionization stages is needed. Hence the dominant source of error for
the abundance will be the uncertainty in the measured line flux and to a
lesser extent the aperture correction. It is very unlikely that the strength
of the silicon line could have been underestimated by a factor of five and it
is even less likely that the aperture correction could be off by such a large
factor. The calcium under-abundance is based on a 3$\sigma$ upper limit for
the [\ion{Ca}{2}] 729.1~nm line. Since Ca$^+$ is a subordinate ionization
stage, the largest source of error will be the correction for unobserved
ionization stages. It seems highly unlikely however that this correction could
be off by a factor of twenty.
 
Recently Liu\markcite{li98} (1998) published an abundance analysis of the
planetary nebula NGC~6153, which yielded abundances derived from recombination
lines that were roughly a factor of 10 higher than those derived from
collisionally excited forbidden lines. Taken at face value, this could cast
doubt on the analysis presented above. Similar discrepancies have been
observed in other nebulae as well, but usually they are much less pronounced.
The strength of this effect for NGC~6445 has not been established. Whether
such a discrepancy exists for magnesium, silicon or calcium has not been
established either since the recombination lines of these lines are too weak
to be observable. The physics causing the discrepancy between collisional and
recombination abundances is not clear and therefore both values should be
viewed with caution until such an explanation is at hand. Liu\markcite{li98}
(1998) claims that there is no evidence that the recombination abundances have
been grossly overestimated. However, the derived O$^{2+}$ abundance of 9.65
for NGC~6153 is much larger than the solar oxygen abundance. Significant
enhancement of oxygen in post-AGB stars is not expected by stellar evolution
theory. We argue that this fact does cast doubt on the recombination
abundances. On the other hand, the O$^{2+}$ abundance derived by
Liu\markcite{li98} (1998) from forbidden lines (using essentially the same
method we use in this paper) is in good agreement with the expected solar
value. Exactly the same argument holds for the Ne$^{2+}$ abundances of
NGC~6153. In view of these facts it seems fair to assume that the collisional
abundances derived in this work are valid. Therefore we claim that there is
evidence for depletion in NGC~6445, which in turn proves that the dust grains
that contain these elements cannot have been destroyed {\sl in situ}, at least
not in significant amounts.

An alternative assumption could be that the ionized region is free of dust
grains due to the fact that the grains were separated from the gas. If
dust-gas separation occurred, it must have been during the post-AGB phase. It
could not have occurred during an earlier phase, since then dust formation
would still be on-going. On the other hand, separation during the early
post-AGB phase could be expected as it is radiation pressure on dust grains
that is thought to drive the outflow of gas and dust at that stage. However,
this phase is very short-lived and it is not likely that a detectable amount
of dust-gas separation could occur in that period. Separation during later
phases seems impossible since our photo-ionization model indicates that once
the nebular material is ionized, the drift velocity with respect to the gas is
lower than 1~km\,s$^{-1}$. This velocity is far too low to produce any
significant dust-gas separation.

We have constructed a model for the grain emission of NGC~6445 based on the
assumption that it contains grains at a constant dust-to-gas ratio throughout
the nebula. Heating of the grains by Ly$\alpha$ photons must be included in
this model. Ly$\alpha$ photons are scattered very efficiently by neutral
hydrogen due to Rayleigh scattering. This process tends to lock up the photons
inside the neutral envelope of the nebula until they are eventually absorbed
by grain particles. However, Ly$\alpha$ photons can escape provided they can
pick up sufficient redshift by repeatedly being bounced off the expanding
neutral envelope. We have included a simplified analytic treatment of this
process in the model. This treatment assumes that there is a sharp boundary
between a fully ionized and a fully neutral region. Hence there is no Rayleigh
scattering inside the ionized region, only outside. It is known that this is
not a good approximation and it will overestimate the escape probability to
some extent. We also included sources of grain heating other than Ly$\alpha$
in our treatment by using the Cloudy model. These calculations indicate that
roughly 25\% of the total grain heating is due to sources other than
Ly$\alpha$ and hence that roughly 25\% of the Ly$\alpha$ photons must escape
in order to account for the observed total grain emission. This shows that
Ly$\alpha$ escape should be included in the model. Given that there are
considerable uncertainties in the shape of the absorption law for the grains
and in the geometry of the nebula, we felt that a more detailed treatment of
Ly$\alpha$ scattering (e.g., using a Monte Carlo code) was not warranted. The
shape of the absorption law is important because the cross section at 121.6~nm
(combined with the dust-to-gas ratio) determines the escape probability, and
because a large infrared emissivity (when compared to the UV absorption) will
tend to make the grains cooler. Also deviations from $\beta = 2$ in the LWS
range will have an important effect on the modeling. None of these parameters
are accurately known for the grains in NGC~6445.

We produced 3 models using single-sized grains of astronomical silicate
(Martin \& Rouleau\markcite{mr91} 1991) with diameters 0.01~$\mu$m, 0.1~$\mu$m
and 1~$\mu$m respectively. We treated the dust-to-gas ratio as a free
parameter, but assumed it to be constant as a function of radius. The results
are shown in Figure~\ref{plot:farir}. The dust-to-gas ratios for the best
fitting models are: 3.0\xt{-3} (0.01~$\mu$m), 4.1\xt{-3} (0.1~$\mu$m) and
1.7\xt{-2} (1~$\mu$m). The models using the 0.01~$\mu$m and 1~$\mu$m grains
are clearly unrealistic, but the 0.1~$\mu$m grains give a reasonable fit. The
sequence of models suggests that the grains in NGC~6445 are on average large.

We mentioned earlier that our model will overestimate the escape probability
of Ly$\alpha$ photons to some extent. Since the total amount of heating of the
grains is constrained by the observations, the model fitted to the
observations will tend to overestimate the dust-to-gas ratio somewhat to
counteract this effect. If we had assumed the silicon-bearing grains to be
crystalline silicates instead of astronomical silicate and if we assume the UV
absorption characteristics of both silicate types to be the same, the
crystalline silicates would have been cooler due to the presence of strong
emission features in the mid- and far-infrared. Under those conditions our
model sequence would have led to a somewhat lower estimate for the average
grain size and dust-to-gas ratio. Using graphite grains would have led to a
model and a dust-to-gas ratio that is roughly similar to what is quoted above.
Given the uncertainties in our model and the absorption and emission
characteristics of the grains in NGC~6445 we cannot derive precise results.
However, it is very likely that the grains in NGC~6445 are large and that the
dust-to-gas ratio is several times $10^{-3}$. One can also predict the
dust-to-gas ratio assuming that the silicon abundance before grain formation
was solar and the grains are composed of pure MgSiFeO$_4$. This yields
3.2\xt{-3}, in excellent agreement with our results. We find there is no need
to invoke dust-gas separation in this nebula in order to obtain a good fit to
the spectrum.

Therefore, the most likely conclusion is that grains are residing inside the
ionized region of NGC~6445 and that the low temperature is caused by the low
luminosity of the central star combined with the large extent of the nebula.
The low flux of the grain emission, when compared to H$\beta$, is caused by
the low optical depth of the grains in the Lyman continuum, which makes
Ly$\alpha$ photons the dominant heating source for the grains. In short, all
of these characteristics are a consequence of the fact that the nebula is well
evolved. Our conclusion implies that the bulk of the silicon-bearing grains in
this nebula could survive exposure to hard UV photons for at least several
thousands of years. This contradicts the results of Natta \&
Panagia\markcite{np81} (1981) and Pottasch\markcite{po87} (1987), who find a
tight correlation between the dust-to-gas ratio and the nebular radius based
on a small sample of PNe. The dust-to-gas ratio in NGC~6445 is much higher
than would be expected from this correlation. The results given by Natta \&
Panagia\markcite{np81} (1981) predict a dust-to-gas ratio of approximately
2\xt{-5} for NGC~6445, i.e. roughly 2 orders of magnitude lower than what we
find. We are confident that our models could not have overestimated the dust
content of the nebula by such a large factor. We cannot rule out that some
destruction of the smallest grains, constituting only a small mass fraction,
could have occurred. Small grains would be the first to be destroyed since
they absorb UV radiation more effectively and are therefore hotter than larger
grains. Such an occurrence would cause the grains in NGC~6445 to be on average
larger. It is not clear whether this is sufficient to explain the large
average grain size discussed above, or if it is simply the natural result of
grain growth (and coagulation) during the prior AGB phase. Natta \&
Panagia\markcite{np81} (1981) found a tight decreasing relation between the
average grain size and the nebular radius based on the same sample mentioned
earlier. This relation would predict an average grains size of 0.007~$\mu$m
for NGC~6445, which is significantly smaller than what we find. Hence this
prediction also seems implausible.

\section{Conclusions}
\label{concl}

\begin{enumerate}
\item
We presented {\it ISO}-SWS and LWS spectra of the planetary nebula NGC~6445
which indicate the presence of very cool dust with a low flux compared to
H$\beta$. The fact that Ly$\alpha$ photons alone have sufficient energy to
account for all of the grain heating, suggests that either the optical depth
of the grains in the Lyman continuum is very low, or that the ionized region
is free of grains.
\item
A photo-ionization model of the ionized region of NGC~6445 was constructed.
Based on this model we showed that it is very unlikely that either grain
destruction or dust-gas separation has taken place in the ionized region.
\item
Hence, the most likely conclusion is that grains reside inside the ionized
region of NGC~6445 and that the low temperature and flux of the grain emission
is caused by the low luminosity of the central star and the low optical depth
of the grains. This implies that the bulk of the silicon-bearing grains in
this nebula could survive exposure to hard UV photons for several thousands of
years, contradicting previously published results.
\item
A comparison between optical and infrared diagnostic line ratios gave a
marginal indication for the presence of a $t^2$-effect. However, the evidence
is not convincing and alternative explanations could be provided by
uncertainties in the absolute flux calibration of the spectra, the aperture
corrections that have been applied or the collisional cross sections.
\item
The photo-ionization model allowed an accurate determination of the central
star temperature using model atmospheres by Rauch\markcite{ra97} (1997). The
resulting value of 184~kK is in good agreement with the average of all
published Zanstra temperatures.
\item
The off-source spectrum taken with LWS clearly shows the presence of a warm
cirrus component with a temperature of 24~K as well as a very cold component
with a temperature of 7~K. Since our observation encompasses only a small
region of the sky, it is not clear how extended the 7~K component is and
whether it contributed significantly to the FIRAS spectrum taken by {\it
COBE}, as reported by Wright et al.\markcite{wr91} (1991) and Reach et
al.\markcite{re95} (1995), but dismissed by Lagache et al.\markcite{la98}
(1998). Because our line of sight is in the galactic plane, the very cold
component could be a starless core.
\end{enumerate}

\acknowledgments
This research was partially based on observations obtained at the European
Southern Observatory and on observations with {\it ISO}, an ESA project with
instruments funded by ESA Member States (especially the PI countries: France,
Germany, the Netherlands and the United Kingdom) and with the participation of
ISAS and NASA. We thank the NSF and NASA for support through grants AST
96-17083 and GSFC--123. We thank Dr.\ A. Lazarian for stimulating discussions
and C. Kerton for processing the IRAS data. The Canadian Institute for
Theoretical Astrophysics is thanked for hospitality and financial support
during our stay in Toronto.

\newpage

\begin{deluxetable}{lllccrccc}
\tablecaption{Log of the Observations. \label{obs:log}}
\scriptsize
\tablehead{
\colhead{} & \colhead{} & \colhead{} & \colhead{Exposure} & \colhead{Aperture\tablenotemark{a}} & \colhead{PA\tablenotemark{b}} & \multicolumn{2}{c}{Central Position} & \colhead{Spectral Range} \nl
\colhead{Telescope} & \colhead{Instrument} & \colhead{Date} & \colhead{(s)} & \colhead{(arcsec$^2$)} & \colhead{(degree)} & \multicolumn{2}{c}{(2000.0)} & \colhead{(\micron)}
}
\startdata
{\it IUE} & SWP17030   & 820524 & 10800    & 9.07$\times$21.65 & 156.9 & 17~49~15.0 & $-$20~00~34 & 0.115 -- 0.198 \nl
          & SWP38267   & 900228 & \phn3600 & 9.07$\times$21.65 & 165.5 & 17~49~14.9 & $-$20~00~26 & 0.115 -- 0.198 \nl
          & LWP17434   & 900227 & \phn3600 & 9.91$\times$22.51 & 165.5 & 17~49~14.9 & $-$20~00~26 & 0.185 -- 0.336 \nl
ESO 1.52m & B\&Ch      & 950516 & \phn\phn120\tablenotemark{c} & 60$\times$140 & 0.0 & 17~49~14.8 & $-$20~00~40 & 0.391 -- 0.968 \nl
{\it ISO} & SWS01      & 970317 & \phn1912\tablenotemark{d} & 20$\times$14 -- 33$\times$20\tablenotemark{e} & 91.1 & 17~49~14.4 & $-$20~00~24 & 2.35 -- 45\phd\phn \nl
          & SWS02      & 970317 & \phn1828\tablenotemark{d} & 20$\times$14 -- 33$\times$20\tablenotemark{e} & 91.1 & 17~49~14.4 & $-$20~00~24 & \nodata\tablenotemark{f} \nl
          & LWS01      & 970317 & \phn\phn640\tablenotemark{d} & \o\ 84\arcsec\ circ.\ & \nodata & 17~49~14.4 & $-$20~00~24 & 43.2 -- 189\phd \nl
          & LWS01 OFF  & 970317 & \phn\phn640\tablenotemark{d} & \o\ 84\arcsec\ circ.\ & \nodata & 17~49~15.3 & $-$19~57~34 & 43.2 -- 189\phd \nl
\enddata
\tablenotetext{a}{The dimensions are EW$\times$NS assuming PA=0\arcdeg.}
\tablenotetext{b}{The aperture rotates counter-clockwise for increasing position angle.}
\tablenotetext{c}{This is the effective exposure time (see \S~\ref{obs:sec}).}
\tablenotetext{d}{This is the Target Dedicated Time.}
\tablenotetext{e}{The aperture size depends on the wavelength (see \S~\ref{aper:corr}).}
\tablenotetext{f}{Selected emission lines were observed (see Table~\ref{flux:tab}).}
\end{deluxetable}

\begin{deluxetable}{llccccc}
\tablecaption{Empirically Determined Values for the Extinction. \label{obs:abs}}
\tablehead{
\colhead{} & \colhead{$\lambda$} & \colhead{Obs.\ Flux} & \colhead{Case B} & \colhead{Pred.\ Flux} & \colhead{1/$\lambda$} & \colhead{$A(\lambda)$} \nl
\colhead{Spectrum} & \colhead{(nm)} & \colhead{(10$^{-14}$ W\,m$^{-2}$)} & \colhead{Ratio} & \colhead{(10$^{-14}$ W\,m$^{-2}$)} & \colhead{(\micron$^{-1}$)} & \colhead{(mag)}
}
\startdata
\ion{H}{1}  & 954.6 & 0.146\phn   &   0.0357   & \phn0.311      &  1.048  &  0.82 \nl
\ion{H}{1}  & 922.9 & 0.0922      &   0.0249   & \phn0.216      &  1.084  &  0.93 \nl
\ion{H}{1}  & 656.3 & 5.19\phn\phn&   2.8158   &    24.5\phn\phn&  1.524  &  1.68 \nl
\ion{He}{2} & 541.2 & 0.0508      &   0.0787   & \phn0.327      &  1.848  &  2.02 \nl
\ion{H}{1}  & 486.1 & 0.762\phn   &   1.0000   & \phn8.70\phn   &  2.057  &  2.64 \nl
\ion{He}{2} & 468.6 & 0.368\phn   &   1.0000   & \phn4.16\phn   &  2.134  &  2.63 \nl
\ion{H}{1}  & 434.0 & 0.334\phn   &   0.4713   & \phn4.10\phn   &  2.304  &  2.72 \nl
\ion{H}{1}  & 410.2 & 0.115\phn   &   0.2612   & \phn2.27\phn   &  2.438  &  3.24 \nl
\ion{H}{1}  & 397.0 & 0.0762      &   0.1605   & \phn1.40\phn   &  2.519  &  3.16 \nl
\ion{He}{2} & 164.0 & 0.399\phn   &   6.9882   &    29.0\phn\phn&  6.098  &  4.65 \nl
\enddata
\end{deluxetable}

\newpage

\begin{deluxetable}{lrrrrl}
\tablewidth{0.9\textwidth}
\tablecaption{The observed line fluxes and dereddened intensities for
NGC~6445. Where appropriate 3~$\sigma$ upper limits are given. \label{flux:tab}}
\tablehead{
\colhead{} & \colhead{$\lambda$} & \colhead{Flux} & \colhead{$\sigma$} & \colhead{Intensity} & \colhead{} \nl
\colhead{Spectrum} & \colhead{(nm/\micron)} & \colhead{(10$^{-14}$~W\,m$^{-2}$)} & \colhead{(percent)} & \colhead{$I({\rm H}\beta) = 100$} & \colhead{Comment}
}
\startdata
 \ion{C}{4}   & 154.9 &        0.015 &    17.1 & \nodata    & IUE1     \nl
              &       &        0.026 &    32.1 & \nodata    & IUE2     \nl
              &       & 0.247\rlap{:\tablenotemark{a}} & 16.2 & 283.\rlap{:\tablenotemark{a}}\phn & combined \nl
 \ion{He}{2}  & 164.0 &        0.024 &    13.7 & \nodata    & IUE1     \nl
              &       &        0.040 &    17.5 & \nodata    & IUE2     \nl
              &       & 0.399\rlap{:\tablenotemark{a}} & 13.2 & 416.\rlap{:\tablenotemark{a}}\phn & combined \nl
 \ion{C}{3}]  & 190.9 &        0.037 &     5.6 & \nodata    & IUE1     \nl
              &       &        0.053 &     8.7 & \nodata    & IUE2     \nl
              &       & 0.556\rlap{:\tablenotemark{a}} &  5.7 & 712.\rlap{:\tablenotemark{a}}\phn & combined \nl
[\ion{Ne}{5}] & 342.6 &        0.050 &    20.0 &       13.9 &       \nl
[\ion{O}{2}]  & 372.6 &        0.838 & \nodata &   209.\phn &       \nl
[\ion{O}{2}]  & 372.9 &        0.769 & \nodata &   192.\phn &       \nl
[\ion{Ne}{3}] & 386.9 &        0.983 & \nodata &   231.\phn &       \nl
[\ion{Ne}{3}] & 396.8 &        0.312 & \nodata &       69.8 &       \nl
 \ion{H}{1}   & 397.1 &        0.076 &    20.3 &       17.0 &       \nl
[\ion{S}{2}]  & 407.2 &        0.055 &    51.2 &       11.6 &       \nl
 \ion{H}{1}   & 410.2 &        0.115 &    19.4 &       23.9 &       \nl
 \ion{H}{1}   & 434.0 &        0.334 &    10.4 &       60.0 &       \nl
[\ion{O}{3}]  & 436.3 &        0.123 &    24.6 &       21.8 &       \nl
 \ion{He}{1}  & 447.1 &        0.058 &    38.7 &        9.6 &       \nl
 \ion{He}{2}  & 468.7 &        0.368 &     6.2 &       53.3 &       \nl
[\ion{Ar}{4}] & 471.3 &        0.044 &    36.2 &        6.2 &       \nl
[\ion{Ar}{4}] & 474.1 &        0.030 &    59.7 &        4.2 &       \nl
 \ion{H}{1}   & 486.1 &        0.762 &     5.0 &   100.\phn &       \nl
[\ion{O}{3}]  & 495.9 &     3.84\phn &     4.6 &   479.\phn &       \nl
[\ion{O}{3}]  & 500.7 & 12.2\phn\phn &     6.5 &  1490.\phn &       \nl
[\ion{N}{1}]  & 520.0 &        0.145 &    18.0 &       16.0 &       \nl
 \ion{He}{2}  & 541.3 &        0.051 &    25.3 &        5.1 &       \nl
[\ion{N}{2}]  & 575.5 &        0.112 &    12.8 &       10.0 &       \nl
 \ion{He}{1}  & 587.6 &        0.178 &     8.2 &       15.3 &       \nl
[\ion{O}{1}]  & 630.0 &        0.503 &     3.6 &       37.9 &       \nl
[\ion{S}{3}]  & 631.2 &        0.037 &    21.2 &        2.8 &       \nl
[\ion{O}{1}]  & 636.4 &        0.153 &     4.6 &       11.3 &       \nl
[\ion{N}{2}]  & 654.8 &     3.19\phn &     8.3 &   224.\phn &       \nl
 \ion{H}{1}   & 656.3 &     5.19\phn &     6.3 &   363.\phn &       \nl
[\ion{N}{2}]  & 658.4 &     9.51\phn &     4.6 &   661.\phn &       \nl
 \ion{He}{1}  & 668.0 &        0.058 &    14.1 &        3.9 &       \nl
[\ion{S}{2}]  & 671.6 &        0.803 &     5.0 &       53.6 &       \nl
[\ion{S}{2}]  & 673.1 &        0.755 &     5.2 &       50.3 &       \nl
[\ion{Ar}{5}] & 700.6 & $<$    0.052 & \nodata & $<$    3.2 &       \nl
 \ion{He}{1}  & 706.5 &        0.056 &     7.7 &        3.4 &       \nl
[\ion{Ar}{3}] & 713.6 &        0.749 &     3.1 &       44.2 &       \nl
[\ion{Ca}{2}] & 729.1 & $<$    0.017 & \nodata & $<$    1.0 &       \nl
[\ion{O}{2}]  & 731.9 &        0.221 &     9.2 &       12.4 &       \nl
[\ion{O}{2}]  & 733.0 &        0.185 &    10.2 &       10.3 &       \nl
[\ion{Ar}{4}] & 733.2 & $<$    0.019 & \nodata & $<$    1.0 &       \nl
[\ion{Ar}{3}] & 775.1 &        0.171 &    12.8 &        8.5 &       \nl
[\ion{Ca}{2}] & 849.8 & $<$    0.033 & \nodata & $<$    1.3 &       \nl
[\ion{Ca}{2}] & 854.2 & $<$    0.035 & \nodata & $<$    1.4 &       \nl
[\ion{S}{3}]  & 906.9 &     1.11\phn &     3.7 &       39.9 &       \nl
 \ion{H}{1}   & 923.2 &        0.092 &    21.1 &        3.2 &       \nl
[\ion{S}{3}]  & 953.2 &     2.73\phn &     7.7 &       91.1 &       \nl
 \ion{H}{1}   & 954.9 &        0.146 &    41.6 &        4.9 &       \nl
 \ion{H}{1}   &   4.1 &        0.094 &     3.5 &        7.2\tablenotemark{a} & SWS02 \nl
[\ion{Mg}{4}] &   4.5 &        0.074 &     8.6 &        5.5\tablenotemark{a} & SWS02 \nl
[\ion{Ar}{6}] &   4.5 & $<$    0.051 & \nodata & $<$    3.8\tablenotemark{a} & SWS02 \nl
[\ion{Mg}{5}] &   5.6 & $<$    0.142 & \nodata & $<$   10.4\tablenotemark{a} & SWS02 \nl
[\ion{Ne}{6}] &   7.6 &        0.052 &    10.3 &        3.8\tablenotemark{a} & SWS02 \nl
[\ion{Ar}{3}] &   9.0 &        0.339 &    14.9 &       27.4\tablenotemark{a} & SWS01 \nl
[\ion{S}{4}]  &  10.5 &        0.954 &     9.7 &       77.0\tablenotemark{a} & SWS01 \nl
[\ion{Ne}{2}] &  12.8 &        0.377 &     1.6 &       24.7\tablenotemark{a} & SWS02 \nl
[\ion{Mg}{5}] &  13.5 & $<$    0.032 & \nodata & $<$    2.1\tablenotemark{a} & SWS02 \nl
[\ion{Ne}{5}] &  14.3 &     1.64\phn &     1.8 &      105.\tablenotemark{a}\phn & SWS02 \nl
[\ion{Ne}{3}] &  15.6 &     4.05\phn &     1.0 &      259.\tablenotemark{a}\phn & SWS02 \nl
[\ion{S}{3}]  &  18.7 &        0.796 &    11.9 &       51.5\tablenotemark{a} & SWS01 \nl
[\ion{Ar}{3}] &  21.8 & $<$    0.276 & \nodata & $<$   17.7\tablenotemark{a} & SWS01 \nl
[\ion{Ne}{5}] &  24.3 &     1.98\phn &     1.0 &      126.\tablenotemark{a}\phn & SWS02 \nl
[\ion{O}{4}]  &  25.9 & 10.8\phn\phn &     1.2 &      686.\tablenotemark{a}\phn & SWS02 \nl
[\ion{S}{3}]  &  33.5 &     1.44\phn &    14.3 &       67.0\tablenotemark{a} & SWS01 \nl
[\ion{Si}{2}] &  34.8 &        0.488 &    26.4 &       22.6\tablenotemark{a} & SWS01 \nl
[\ion{Ne}{3}] &  36.0 &        0.360 &     5.4 &       16.7\tablenotemark{a} & SWS02 \nl
[\ion{O}{3}]  &  51.7 & 19.0\phn\phn &     1.8 &   255.\phn &       \nl
[\ion{N}{3}]  &  57.3 &     6.30\phn &     1.6 &       84.4 &       \nl
[\ion{O}{1}]  &  63.2\tablenotemark{b} & 4.71\phn & 1.4 & 63.0 &     \nl
              &       &        0.124 &    23.4 &        1.7 & off source \nl
[\ion{O}{3}]  &  88.2 & 10.9\phn\phn &     1.4 &   146.\phn &       \nl
[\ion{N}{2}]  & 121.8 &        0.402 &     2.9 &        5.4 &       \nl
[\ion{O}{1}]  & 145.6\tablenotemark{b} & 0.235 & 2.3 &  3.1 &      \nl
[\ion{C}{2}]  & 157.7 &        0.894 &     1.7 &       12.0\tablenotemark{c} &       \nl
              &       &        0.241 &    14.0 &        3.2\tablenotemark{c} & off source \nl
\enddata
\tablenotetext{a}{An aperture correction factor has been applied.}
\tablenotetext{b}{This line was not used to constrain the photo-ionization model.}
\tablenotetext{c}{The off-source flux was subtracted from the on-source flux to
constrain the modeling.}
\end{deluxetable}

\begin{deluxetable}{lrrl}
\tablewidth{0.8\textwidth}
\tablecaption{Plasma diagnostics for NGC~6445, using various line ratios.
\label{plasma:diag}}
\tablehead{
 & & \colhead{$T_{\rm e}$} & \colhead{$n_{\rm e}$} \nl
Diagnostic & Ratio & \colhead{(kK)} & \colhead{(cm$^{-3}$)}
}
\startdata
[\ion{N}{2}] (654.8+658.4)/575.5 & 88$\pm$12       & 10.4$\pm$0.6 & \nl
[\ion{S}{3}] (906.9+953.2)/631.2 & 48$\pm$10       & 11.5$\pm$1.1 & \nl
[\ion{O}{3}] (495.9+500.7)/436.3 & 90$\pm$23       & 13.5$\pm$1.2 & \nl
[\ion{S}{3}] 953.2/18.7          & 1.8$\pm$0.5     &  8.9$\pm$1.6 & \nl
[\ion{O}{3}] 500.7/51.9          & 5.8$\pm$1.2     & 11.5$\pm$0.9 & \nl
[\ion{Ne}{3}] 386.9/15.6         & 0.89$\pm$0.26   & 11.0$\pm$0.9 & \nl
[\ion{Ne}{5}] 342.6/14.3         & 0.13$\pm$0.04   & 12.0$\pm$0.8 & \nl
[\ion{S}{2}] 671.6/673.1         & 1.07$\pm$0.08   & & 470$\pm$170 \nl
[\ion{O}{2}] 372.6/372.9         & 1.09$\pm$0.11   & & 500$\pm$150 \nl
[\ion{S}{3}] 18.7/33.5           & 0.77$\pm$0.14   & & 310$\pm$200 \nl
[\ion{O}{3}] 51.9/88.2           & 1.75$\pm$0.18   & & 550$\pm$90 \nl
\enddata
\end{deluxetable}

\begin{deluxetable}{cccc}
\tablewidth{0.6\textwidth}
\tablecaption{Parameters for the fits to the far-infrared emission
(45~\micron\ -- 190~\micron) of the nebular grains and ISM grains
respectively.
\label{farir:tab}}
\tablehead{
\multicolumn{4}{c}{Nebular grains} \nl
 & \colhead{$T_{\rm dust}$} & \colhead{FIR} & \nl
\colhead{$\beta$} & \colhead{(K)} & \colhead{(10$^{-14}$~W\,m$^{-2}$)} & \colhead{$\chi^2/N$}
}
\startdata
0.00 & 84.6 & 301 & 2.21 \nl
0.50 & 66.5 & 259 & 1.59 \nl
0.89 & 57.3 & 238 & 1.45 \nl
1.00 & 55.2 & 233 & 1.46 \nl
1.50 & 47.3 & 215 & 1.74 \nl
2.00 & 41.5 & 201 & 2.32 \nl
\nl
\multicolumn{4}{c}{ISM background} \nl
 & \colhead{$T_{\rm warm}$} & \colhead{$T_{\rm cold}$} & \nl
\colhead{$\beta$} & \colhead{(K)} & \colhead{(K)} & \colhead{$\chi^2/N$} \nl
\nl
2.00 & 24.4 & 6.7 & 0.92 \nl
\enddata
\end{deluxetable}

\begin{deluxetable}{lrr}
\tablewidth{0.7\textwidth}
\tablecaption{Parameters for the {\sc cloudy} model of NGC~6445. All symbols
have their usual meaning.
\label{mod:par}}
\tablehead{
\colhead{Parameter} & \colhead{Value} & \colhead{rel.\ to solar\tablenotemark{a}}
}
%
%
\startdata
$T_{\rm eff}$ (kK)            & 184.     & \nl
$L_{\ast}$ (L$_{\odot}$)      & 1035.    & \nl
$r_{\rm in}$ (mpc)            & 79.      & \nl
$r_{\rm Str}$ (mpc)           & 183.     & \nl 
$n_{\rm H}$ (cm$^{-3}$)       & 525\tablenotemark{b} & \nl
$T_{\rm e}$ (kK)              & 11.49\tablenotemark{c} & \nl
$n_{\rm e}$ (cm$^{-3}$)       & 571\tablenotemark{c} & \nl
$\epsilon$(He)\tablenotemark{d} & 11.18  & $+$0.19 \nl
$\epsilon$(C)                 &  8.87\rlap{:} & $+$0.32\rlap{:} \nl
$\epsilon$(N)                 &  8.39    & $+$0.42 \nl
$\epsilon$(O)                 &  8.87    & $+$0.00 \nl 
$\epsilon$(Ne)                &  8.31    & $+$0.23 \nl
$\epsilon$(Mg)                &  7.23\rlap{:} & $-$0.35\rlap{:} \nl
$\epsilon$(Si)                &  6.86    & $-$0.69 \nl
$\epsilon$(S)                 &  6.90    & $-$0.43 \nl   
$\epsilon$(Ar)                &  6.58    & $+$0.06 \nl
$\epsilon$(Ca)                & $<$ 5.09 & $< -$1.27 \nl
$D$ (kpc)                     & 1.5\tablenotemark{b,e} & \nl
$\chi^2$                      &   2.9    & \nl 
\enddata
\tablenotetext{a}{Grevesse, Noels \& Sauval\markcite{gr96} (1996).}
\tablenotetext{b}{This parameter was not varied.}
\tablenotetext{c}{This is a weighted average over the entire nebula.}
\tablenotetext{d}{$\epsilon(A)$ stands for the logarithmic abundance
of $A$; $\epsilon$(H) $\equiv 12$.}
\tablenotetext{e}{Van de Steene \& Zijlstra\markcite{st95} (1995).}
\end{deluxetable}

\begin{deluxetable}{cccr}
\tablecaption{Various determinations of the stellar temperature.
\label{cst:tab}}
\tablehead{
\colhead{$T_{\rm Z}$(\ion{H}{1})} & \colhead{$T_{\rm Z}$(\ion{He}{2})} & \colhead{Other Methods} & \nl
\colhead{(kK)} & \colhead{(kK)} & \colhead{(kK)} & \colhead{Ref}
}
\startdata
       & 182.  &           &  1 \nl
       &       & 174. (ZM) &  2 \nl
       & 184.  &           &  3 \nl
 250.  & 255.  &           &  4 \nl
 184.  & 186.  &           &  5 \nl
       &       & 145. (EB) &  6 \nl
       &       & 180. (He) &  7 \nl
 175.  & 175.  &           &  8 \nl
 162.  & 166.  &           &  9 \nl
 224.  & 186.  &           & 10 \nl
       &       & 172. (He) & 11 \nl
 183.  & 215.  &           & 12 \nl
       &       & 186. (ZA) & 13 \nl
       &       & 184. (ZC) & 14 \nl
 177.  & 180.  &           & 15 \nl
       &       & 194. (EB) & 16 \nl
 182.  & 214.  & 182. (ZC) & 17 \nl
 178.  & 182.  &           & 17 \nl
 174.  & 159.  & 166. (EB) & 18 \nl
\enddata
\tablenotetext{}{References ---
1. Harman \& Seaton\markcite{hs66} 1966,\x{2}
2. Pilyugin, Sakhibullin \& Khromov\markcite{pi78} 1978,\x{2}
3. K\"oppen \& Tarafdar\markcite{kt78} 1978,\x{2}
4. Pottasch\markcite{po81} 1981,\x{2}
5. Kaler\markcite{ka83} 1983 (using a covering factor $\xi = 1.00$.),\x{2}
6. Preite-Martinez \& Pottasch\markcite{pp83} 1983,\x{2}
7. Che \& K\"oppen\markcite{ck83} 1983,\x{2}
8. Reay et al.\markcite{re84} 1984,\x{2}
9. Tylenda\markcite{ty86} 1986,\x{2}
10. Sabbadin\markcite{sb86} 1986b,\x{2}
11. Gurzadyan\markcite{gu88} 1988,\x{2}
12. Gathier \& Pottasch\markcite{gp88} 1988,\x{2}
13. Tylenda\markcite{ty89} 1989,\x{2}
14. Kaler \& Jacoby\markcite{kj89} 1989,\x{2}
15. Jacoby \& Kaler\markcite{jk89} 1989,\x{2}
16. Preite-Martinez et al.\markcite{pr89} 1989,\x{2}
17. Mallik\markcite{ma91} 1991,\x{2}
18. Mal'kov, Golovatyj \& Rokach\markcite{ma95} 1995.
}
\tablenotetext{}{Methods ---
EB: Energy Balance or Stoy method,
He: Empirical relation between $I(468.6~\nm)$ and T$_\ast$,
ZA: Average of hydrogen and helium Zanstra temperature,
ZC: Zanstra temperature using cross-over technique,
ZM: Helium Zanstra temperature method using atmosphere models.
}
\end{deluxetable}

\begin{deluxetable}{lrrrrrr}
\tablewidth{0.8\textwidth}
\tablecaption{Predicted and observed central star magnitudes.
\label{mstar:tab}}
\tablehead{
 & \colhead{\sl U} & \colhead{\sl B} & \colhead{479.3\tablenotemark{a}} & \colhead{\sl V} & \colhead{\sl R} & \colhead{\sl I}
}
\startdata
model              & 14.82 & 16.30 & 16.43 & 16.66 & 16.85 & 17.10 \nl
$A(\lambda)$ (mag) &  3.24 &  2.74 &  2.52 &  2.14 &  1.68 &  1.14 \nl
predicted          & 18.06 & 19.04 & 18.95 & 18.80 & 18.53 & 18.24 \nl
observed           &       &       & 18.97\tablenotemark{b} & 18.9\tablenotemark{c}\phn &       &       \nl
                   &       &       &       & 19.04\tablenotemark{d} &       &       \nl
\enddata
\tablenotetext{a}{ESO 1402 filter centered on 479.3~nm}
\tablenotetext{b}{Gathier \& Pottasch\markcite{gp88} 1988}
\tablenotetext{c}{Reay et al.\markcite{re84} 1984}
\tablenotetext{d}{Jacoby \& Kaler\markcite{jk89} 1989}
\end{deluxetable}

\newpage

\newpage

\begin{references}
\reference{al73} Aller, L. H., Czyzak, S. J., Craine, E., and Kaler J. B. 1973, \apj, 182, 509
\reference{be96} Beintema, D. A., van Hoof, P. A. M., Lahuis, F., Pottasch, S. R., Waters, L. B. F. M., de Graauw, T., Boxhoorn, D. R., Feuchtgruber, H., and Morris, P. W., 1996, \aap, 315, L253
\reference{be98} Bell, K. L., Berrington, K. A., and Thomas, M. R. J. 1998, \mnras, 293, L83
\reference{be92} Bernard, J. P., Boulanger, F., Desert, F. X., and Puget, J. L. 1992, \aap, 263, 258
\reference{bk95} Bhatia, A. K., and Kastner, S. O. 1995, \apjs, 96, 325
\reference{bl95a} Bl\"ocker T. 1995a, \aap, 297, 727
\reference{bl95b} Bl\"ocker T. 1995b, \aap, 299, 755
\reference{bo98} Bottorff, M., LaMothe, J., Momjian, E., Verner, E., Vinkovi\'c, D., and Ferland, G. 1998, \pasp, 110, 1040
\reference{bz94} Butler, K., and Zeippen, C. J. 1994, \aaps, 108, 1
\reference{ca89} Cardelli, J. A., Clayton, C., and Mathis, J. S. 1989, \apj, 345, 245
\reference{ck83} Che, A., and K\"oppen, J. 1983, \aap, 118, 107
\reference{cl96} Clegg, P. E., Ade, P. A. R., Armand, C., et al.\ 1996, \aap, 315, L38
\reference{cs95} Corradi, R. L. M., and Schwarz, H. E. 1995, \aap, 293, 871
\reference{dg96} de Graauw, Th., Haser, L. N., Beintema, D. A., et al.\ 1996, \aap, 315, L49
\reference{di85} Dinerstein, H. L., Lester, D. F., and Werner, M. W. 1985, \apj, 291, 561
\reference{dl84} Draine, B. T., and Lee, H. M. 1984, \apj, 285, 89
\reference{fe98} Ferland, G. J., Korista, K. T., Verner, D. A., Ferguson, J. W., Kingdon, J. B., and Verner, E. M. 1998, \pasp, 110, 761
\reference{ga95} Galavis, M. E., Mendoza, C., and Zeippen, C. J. 1995, \aaps, 111, 347
\reference{ga97} Galavis, M. E., Mendoza, C., and Zeippen, C. J. 1997, \aaps, 123, 159
\reference{gp88} Gathier, R., and Pottasch, S. R. 1988, \aap, 197, 266
\reference{gr96} Grevesse, N., Noels, A., and Sauval, A. J. 1996, in
`Cosmic Abundances', eds.\ Holt, S. S., and Sonneborn, G., ASP Conference
series, Vol.\ 99, p.~117
\reference{gu88} Gurzadyan, G. A. 1988, \apss, 149, 343
\reference{hs66} Harman, R. J., and Seaton, M. J. 1966, \mnras, 132, 15
\reference{ho99} Hotzel, S., Lemke, D., T\'oth, L. V., et al.\ 1999, `The Universe as seen by ISO', eds.\ Cox, P., and Kessler, M. F., ESA Publication SP-427, Noordwijk, p.~675
\reference{hh89} Huggins, P. J., and Healy, A. P. 1989, \apj, 346, 201
\reference{jk89} Jacoby, G. H., and Kaler J. B. 1989, \aj, 98, 1662
\reference{ka83} Kaler, J. B. 1983, \apj, 271, 188
\reference{kj89} Kaler, J. B., and Jacoby, G. H. 1989, \apj, 345, 871
\reference{km95} Kim, S.-H., and Martin, P. G. 1995, \apj, 442, 172
\reference{ki97} Kingdon, J. B., and Ferland, G. J. 1997, \apj, 477, 732
\reference{ki95} Kingdon, J., Ferland, G. J., and Feibelman, W. A. 1995, \apj, 439, 793
\reference{kt78} K\"oppen, J., and Tarafdar, S. P. 1978, \aap, 69, 363
\reference{la98} Lagache, G., Abergel, A., Boulanger, F., and Puget, J.-L. 1998, \aap, 333, 709
\reference{lr77} Launay, J. M., and Roueff, E. 1977, \aap, 56, 289
\reference{la99} Laureijs, R. J. 1999, `The Universe as seen by ISO', eds.\ Cox, P., and Kessler, M. F., ESA Publication SP-427, Noordwijk, p.~599
\reference{lb94} Lennon, D. J., and Burke, V. M. 1994, \aaps, 103, 273
\reference{le89} Lenzuni, P., Natta, A., and Panagia, N. 1989, \apj, 345, 306
\reference{li98} Liu, X.-W. 1998, \apss, 255, 499
\reference{ma95} Mal'kov, Y. F., Golovatyj, V. V., and Rokach, O. V. 1995, \apss, 232, 99
\reference{ma91} Mallik, D. C. V. 1991, Publ.\ Astr.\ Soc.\ Austr., 9, 15
\reference{mr91} Martin, P. G., and Rouleau, F. 1991, in `Extreme Ultraviolet Astronomy', eds.\ Malina, R. F., and Bowyer, S., Pergamon Press, New York, p.~341
\reference{ma90} Mathis, J. S. 1990, \araa, 28, 37
\reference{mz82} Mendoza, C., and Zeippen, C. J. 1982, \mnras, 199, 1025
\reference{ma82} Milne, D. K., and Aller, L. H. 1982, \aaps, 50, 209
\reference{mf87} Monteiro, T. S., and Flower, D. R. 1987, \mnras, 228, 101
\reference{np81} Natta, A., and Panagia, N. 1981, \apj, 248, 189
\reference{od63} O'Dell, C. R. 1963, \apj, 138, 293
\reference{ol96} Oliva, E., Pasquali, A., and Reconditi, M. 1996, \aap, 305, L21
\reference{pe67} Peimbert, M. 1967, \apj, 150, 825
\reference{pe91} Perinotto, M. 1991, \apjs, 76, 687
\reference{pm88} Phillips, J. P., and Mampaso, A. 1988, \aap, 190, 237
\reference{pi78} Pilyugin, L. S., Sakhibullin, N. A., and Khromov, G. S. 1978, Astrophysics, 14, 377
\reference{po81} Pottasch, S. R. 1981, \aap, 94, L13
\reference{po84} Pottasch, S. R. 1984, `Planetary Nebulae', Reidel, Dordrecht
\reference{po87} Pottasch, S. R. 1987, in `Late Stages of Stellar Evolution', eds.\ Kwok, S., and Pottasch, S. R., Reidel, Dordrecht, p.~355
\reference{pp83} Preite-Martinez, A., and Pottasch, S. R. 1983, \aap, 126, 31
\reference{pr89} Preite-Martinez, A., Acker, A., K\"oppen, J., and Stenholm, B. 1989, \aaps, 81, 309
\reference{ra97} Rauch, T. 1997, \aap, 320, 237
\reference{re95} Reach, W. T., Dwek, E., Fixsen, D. J. et al. 1995, \apj, 451, 188
\reference{re84} Reay, N. K., Pottasch, S. R., Atherton, P. D., and Taylor, K. 1984, \aap, 137, 113
\reference{rl85} Rieke, G. H., and Lebofsky, M. J. 1985, \apj, 288, 618
\reference{ro93} Rowlands, N., Houck, J. R., Skrutskie, M. F., and Shure, M. 1993, \pasp, 105, 1287
\reference{sa86} Sabbadin, F. 1986a, \aaps, 64, 579
\reference{sb86} Sabbadin, F. 1986b, \aap, 160, 31
\reference{sc92} Schwarz, H. E., Corradi, R. L. M., and Melnick, J. 1992, \aaps, 96, 23
\reference{sd95} Shaw, R. A., and Dufour, R. J. 1995, \pasp, 107, 896
\reference{si75} Simpson, J. P. 1975, \aap, 39, 43
\reference{sh95} Storey, P. J., and Hummer, D. G. 1995, \mnras, 272, 41
\reference{to99} T\'oth, L. V., Lemke, D., Krause O., et al.\ 1999, `The Universe as seen by ISO', eds.\ Cox, P., and Kessler, M. F., ESA Publication SP-427, Noordwijk, p.~771
\reference{ty86} Tylenda, R. 1986, \aap, 156, 217
\reference{ty89} Tylenda, R. 1989, in Proc.\ IAU Symp.\ 131: `Planetary Nebulae', ed.\ Torres-Peimbert, S., Kluwer, Dordrecht, p.~531
\reference{st95} Van de Steene, G. C., and Zijlstra, A. A. 1995, \aap, 293, 541
\reference{vv99} van Hoof, P. A. M., and Van de Steene, G. C. 1999, \mnras, submitted
\reference{vh97a} van Hoof, P. A. M., Oudmaijer, R. D., and Waters L. B. F. M. 1997a, \mnras, 289, 371
\reference{vh97b} van Hoof, P. A. M., Verner, D. A., Beintema, D. A., and
Ferland, G. J. 1997b, in `Proceedings of the First {\it ISO} Workshop on Analytical Spectroscopy', eds.\ Heras, A. M., Leech, K., Trams, N. R., and Perry, M.,
ESA Publication SP-419, Noordwijk, p.~235
\reference{vh99} van Hoof, P. A. M., Beintema, D. A., Verner, D. A., and
Ferland, G. J. 1999, \aap, submitted
\reference{wt94} Ward-Thompson, D., Scott, P. F., Hills, R. E., and Andr\'e, P. 1994, \mnras, 268, 276
\reference{wa96} Waters, L. B. F. M., Molster F. J., de Jong, T., et al.\ 1996, \aap, 315, L361
\reference{ws80} Watson, D. M., and Storey, J. W. V. 1980, Int.\ J.\ Infrared Millimeter Waves, 1, 60
\reference{we89} Weinberger, R. 1989, \aaps, 78, 301
\reference{wr91} Wright, E. L., Mather, J. C., Bennett, C. L. et al. 1991, \apj, 381, 200
\reference{zk93} Zhang, C. Y., and Kwok, S. 1993, \apjs, 88, 137
\end{references}
\end{document}